\documentclass[prx,preprint,showpacs,preprintnumbers,amsmath,amssymb,floatfix,dvipdfmx]{revtex4-1}
\usepackage{graphicx}

\usepackage{bm}% bold math
\usepackage{xcolor}
\usepackage{url}
\usepackage{bookmark}
\newcommand{\be}{\begin{equation}}
\newcommand{\ee}{\end{equation}}
\newcommand{\eq}[1]{Eq.~(\ref{#1})}
\newcommand{\fig}[1]{Fig.~\ref{#1}}
\def\bea{\begin{eqnarray}}
\def\eea{\end{eqnarray}}

\def\vq{{\bf q}}

\def\vk{{\bf k}}
\def\vQ{{\bf Q}}

\begin{document}

\title{Retaining Landau quasiparticles in the presence of realistic charge fluctuations in cuprates} 

\author{Hiroyuki Yamase$^{1}$, Mat\'{\i}as Bejas$^{2}$, and Andr\'es Greco$^{2}$}
\affiliation{
{$^1$}Research Center for Materials Nanoarchitectonics (MANA), National Institute for Materials Science (NIMS), Tsukuba 305-0047, Japan\\
{$^2$}Facultad de Ciencias Exactas, Ingenier\'{\i}a y Agrimensura and Instituto de F\'{\i}sica Rosario (UNR-CONICET), Avenida Pellegrini 250, 2000 Rosario, Argentina
}

%\date{\today}
\date{October 26, 2023}

\begin{abstract}
Charge excitation spectra are getting clear in cuprate superconductors in momentum-energy space especially around a small momentum region, where plasmon excitations become dominant. Here, we study whether  Landau quasiparticles survive in the presence of charge fluctuations observed in experiments. We employ the layered $t$-$J$ model with the long-range Coulomb interaction, which can reproduce the realistic charge fluctuations. We find that Landau quasiparticles are retained in a realistic temperature and doping region, although the quasiparticle spectral weight is strongly reduced to $0.08$-$0.24$. Counterintuitively, the presence of this small quasiparticle weight does not work favorably to generate a pseudogap. 
\end{abstract}

%\pacs{74.20.Mn, 75.25.Dk, 74.20.Rp, 74.70.Xa}

\maketitle

\section{introduction}
Angle-resolved photoemission spectroscopy (ARPES) is a powerful tool to reveal the one-particle excitation spectrum \cite{damascelli03}. In cuprate high-temperature superconductors, it shows a gaplike feature---the spectral weight at Fermi momenta is suppressed at zero energy already well above the superconducting transition temperature $T_{c}$, leading to a peak at a finite energy. This phenomenon is known as the pseudogap \cite{timusk99,keimer15}. It develops already around the optimally doped region and is pronounced  in the underdoped region. Since the high temperature superconductivity is realized inside the pseudogap phase, the understanding of the pseudogap is one of the most important issues in the cuprate phenomenology and has been studied intensively. Despite many studies, however, the pseudogap is still a controversial issue and remains elusive. 

Recently, resonant inelastic x-ray scattering (RIXS) revealed charge excitation spectra in momentum-energy space especially around the zone center in both electron-doped \cite{hepting18,lin20,hepting22} and hole-doped  \cite{nag20,hepting22,singh22a} cuprates. They were identified as plasmon excitations specific to layered metallic systems---not only the usual optical plasmon but also acoustic-like plasmon modes are present \cite{greco16}. Given that the plasmon energies are low \cite{nag20,hepting22,singh22a}, it is important to examine the role of realistic three-dimensional charge fluctuations in the low-energy quasiparticle properties, including the pseudogap phenomenology. 

Nonetheless, many theoretical studies were performed not only in two-dimensional models but also for a short-range interaction---hence there are no plasmons. Reference~\cite{dong19} concluded that charge fluctuations do not lead to a pseudogap in the dynamical cluster approximation to the two-dimensional Hubbard model. This conclusion is shared with Refs.~\cite{gunnarsson15,schafer21}, where the pseudogap is associated with antiferromagnetic fluctuations.  However, charge fluctuations considered in Ref.~[\onlinecite{dong19}] are qualitatively different from the actual experimental data. 
Moreover, a recent work in the dynamical cluster approximation indicates that antiferromagnetic fluctuations alone cannot capture the pseudogap \cite{yu24}. 

The situation is also similar in research of a strange metal physics, which currently attracts much interest especially in the context of Planckian dissipation in metals \cite{patel19,grissonnanche21,phillips22}. Recent experiments \cite{mitrano18,husain19,arpaia23} discussed that charge fluctuations can be responsible for the strange metallic properties in cuprates. Theoretical studies in Refs.~\cite{seibold21,caprara22} are in line with this scenario. However, they missed realistic three-dimensional charge excitations including plasmons.

Realistic charge excitation spectra were reproduced in a large-$N$ theory of the $t$-$J$ model with the long-range Coulomb interaction \cite{greco19,greco20,nag20,hepting22}---we may refer to it as the $t$-$J$-$V$ model \cite{greco16}. Two theoretical studies were performed about the electron self-energy in the $t$-$J$-$V$ model. First, plasmon excitations were found to generate a fermionic incoherent  band---{\it plasmaron} dispersion---near the energy of the optical plasmon energy \cite{yamase23}. Second, quantum charge fluctuations, namely fluctuations at zero temperature, lead to a side band with an energy scale higher than the plasmarons, but on the opposite energy side across the Fermi energy \cite{yamase21a}. Considering that their energy scale is high, these features may not depend on temperature, which validates calculations at zero temperature in Refs.~\cite{yamase21a,yamase23}. However, both pseudogap and Planckian dissipation are related to not only finite temperature but also a  low-energy property of electrons close to the Fermi surface. These phenomena are associated with the charge degree of freedom of electrons, suggesting a possibly pivotal role of charge fluctuations. 

In this paper, we achieve accurate numerics even in a low-energy region at finite temperatures in the large-$N$ theory of the $t$-$J$-$V$ model. This technical success allows us to study closely the electron self-energy by taking the realistic three-dimensional charge fluctuations into account. While the system loses approximately 75-90 \% of the quasiparticle weight on the entire Fermi surface,  we find Landau quasiparticles in a realistic temperature and doping region. This implies that the charge fluctuations are not responsible for 
the pseudogap and a strange metal physics. We also find that the small quasiparticle weight is not effective to generate a pseudogap even if additional self-energy corrections are considered, implying an intriguing role of charge fluctuations in the pseudogap state.

\section{Formalism} 
The $t$-$J$ model is a microscopic model of cuprate superconductors \cite{anderson87,fczhang88,lee06}. To capture the plasmon physics in cuprates and achieve realistic calculations, we employ the following layered $t$-$J$-$V$ model: 

\begin{equation}
H = -\sum_{i, j,\sigma} t_{i j}\tilde{c}^\dag_{i\sigma}\tilde{c}_{j\sigma} + 
\sum_{\langle i,j \rangle} J_{ij} \left( \vec{S}_i \cdot \vec{S}_j - \frac{1}{4} n_i n_j \right)
+  \frac{1}{2} \sum_{i \neq j} V_{ij} n_i n_j \,.
\label{tJV}  
\end{equation}
Here $\tilde{c}^\dag_{i\sigma}$ ($\tilde{c}_{i\sigma}$) are the creation (annihilation) operators of electrons with spin $\sigma (=\uparrow, \downarrow)$  in the Fock space without double occupancy at any site---strong correlation effects,  $n_i=\sum_{\sigma} \tilde{c}^\dag_{i\sigma}\tilde{c}_{i\sigma}$ is the electron density operator, and $\vec{S}_i$ is the spin operator. The sites $i$ and $j$ run over a layered square lattice. The hopping $t_{i j}$ takes the value $t$ $(t')$ between the first (second) nearest-neighbor sites on the square lattice and is scaled by $t_z$ between the layers. The exchange interaction $J_{i j}=J$ is considered only for the nearest-neighbor sites in the layer as denoted by $\langle i,j \rangle$---the exchange term between the layers is much smaller than $J$ (Ref.~[\onlinecite{thio88}]). $V_{ij}$ is the long-range Coulomb interaction and  describes plasmon excitations. The layered structure is a requisite to describe not only the usual optical plasmon but also acoustic-like plasmons \cite{grecu73,fetter74,grecu75}.

In momentum space $V_{ij}$ is written as \cite{becca96} 
\be
V(\vq)=\frac{V_c}{\alpha (2 - \cos q_x - \cos q_y)+1 - \cos q_z} \,,
\label{LRC}
\ee
where $V_c= e^2 d(2 \epsilon_{\perp} a^2)^{-1}$ and $\alpha=\frac{\epsilon_\parallel/\epsilon_\perp}{(a/d)^2}$; $e$ is the electric charge of electrons, $a$ the unit length of the square lattice, $d$ the distance between the layers, and $\epsilon_\parallel$ and $\epsilon_\perp$ are the dielectric constants parallel and perpendicular to the planes, respectively. In \eq{LRC}, $\alpha$ describes the anisotropy between the in-plane and out-of-plane interaction.

We analyze the model (\ref{tJV}) by using a large-$N$ technique in a path integral representation in terms of the Hubbard operators \cite{foussats04}.  In this scheme, charge fluctuations associated with the usual charge-density-wave and plasmons are described  by a $2 \times 2$ matrix $D_{ab}(\vq, i \nu_{n})$ with $a,b=1,2$; $\vq$ is  the momentum of the charge fluctuations and $\nu_{n}$ a bosonic Matsubara frequency. While $D_{11}$ corresponds to the usual density-density correlation function, $D_{22}$ is a special feature of strong correlation effects---it describes fluctuations associated with the local constraint. Naturally the off-diagonal component $D_{12} (=D_{21})$ is also present. Strictly speaking, there are also bond-charge fluctuations, which can be incorporated by enlarging $D_{ab}$ to a $6 \times 6$ matrix. The bond-charge fluctuations are, however, less effective on the electron self-energy than the usual charge fluctuations \cite{yamase21a} and are thus neglected for simplicity. After the analytical continuation $i \nu_{n}\rightarrow \nu + i \Gamma_{\rm ch}$, where $\Gamma_{\rm ch} (>0)$ is infinitesimally small, we obtain the full charge excitation spectrum described by ${\rm Im}D_{ab}(\vq, \nu)$, which contains both plasmon excitations as well as gapless particle-hole excitations---see Ref.~[\onlinecite{bejas17}] for a comprehensive analysis of ${\rm Im}D_{ab}(\vq, \nu)$.

The charge fluctuations can renormalize the one-particle property of electrons, which can be analyzed by computing the electron self-energy. This requires involved calculations in the large-$N$ theory because one needs to go beyond leading order theory. At order of $1/N$, the imaginary part of the self-energy is calculated  as \cite{yamase21a} 
\be
{\mathrm{Im}}\Sigma_{\rm ch} ({\mathbf{k}},\omega)= \frac{-1}{N_{s} N_z}
\sum_{a,b}\sum_{{\mathbf{q}}} {\rm Im}D_{ab} (\vq,\nu) h_{a}(\vk,\vq,\nu) 
h_{b}(\vk,\vq,\nu) \left[
n_{F}( -\varepsilon_{\vk-\vq}) +n_{B}(\nu) 
\right] \, .
\label{ImSig}
\ee
Here $\nu = \omega -\varepsilon_{\vk - \vq}$,  $\varepsilon_{\vk}$ is the electron dispersion obtained at leading order, $h_{a}(\vk, \vq, \nu)$ a vertex describing the coupling between electrons and charge excitations, $n_{F}$ and $n_{B}$ the Fermi and Bose distribution functions, respectively, $N_{s}$ the total number of lattice sites in each layer, and $N_{z}$ the number of layers; see Ref.~\cite{yamase21a} for the explicit forms of $D_{ab} (\vq, \nu)$, $\varepsilon_{\vk}$, and $h_{a}(\vk, \vq, \nu)$.  The above self-energy has the same structure as a self-energy obtained from the Fock diagram in a perturbation theory. However, in the large-$N$ scheme, we have both Hartree and Fock diagrams in a nontrivial way at order of $1/N$. Moreover, it includes charge fluctuations associated with the local constraint described by ${\rm Im}D_{12}$ and ${\rm Im}D_{22}$.

The real part of $\Sigma_{\rm ch} (\vk, \omega)$ is calculated by the Kramers-Kronig relations. Since the electron Green's function $G(\vk,\omega)$ is written as $G^{-1}(\vk,\omega) = \omega + \mathrm{i} \Gamma_{\rm sf} -\varepsilon_{\vk} - \Sigma_{\rm ch} (\vk,\omega)$, we obtain the one-particle spectral function $A({\bf k},\omega)=-\frac{1}{\pi} {\rm Im}G({\bf k},\omega)$: 
\be
A({\bf k},\omega)= -\frac{1}{\pi} \frac{{\rm Im}\Sigma_{\rm ch}({\bf k},\omega) - \Gamma_{\rm sf}}
{[\omega- \varepsilon_{\vk}-{\rm Re}\Sigma_{\rm ch}({\bf k},\omega)]^2 
+ [{\rm Im}\Sigma_{\rm ch}(\vk,\omega) -\Gamma_{\rm sf}]^2} \,, 
\label{Akw}
\ee
where $\Gamma_{\rm sf} (>0)$ originates from the analytical continuation in the electron Green's function.

\section{Results}
\subsection{Role of realistic charge fluctuations} 
We choose parameters $t'/t=-0.20$, $J/t=0.3$, $t_{z}/t=0.01$, $V_c/t=31$, $\alpha=3.5$, $\Gamma_{\rm ch}=\Gamma_{\rm sf}=0.03t$, and $N_{z}=10$, and put $t=1$ as the energy unit. These parameters were obtained to describe the plasmon dispersion  observed in ${\rm La_{2-x}Sr_xCuO_4}$ (LSCO) \cite{hepting22}. In LSCO, the plasmon energy with a finite $q_z$ becomes less than 55 meV  at the in-plane zone center  \cite{hepting22}. This low-energy plasmon seems to offer a favorable situation where plasmon excitations could affect effectively the electron property around the Fermi surface. We thus focus on a small energy window around $\omega=0$ in this paper. Because of the layered model, the Fermi surface depends on $k_{z}$. Our conclusions, however, do not depend on $k_{z}$ and we have presented results for $k_{z}=0$.

%%%%%%%%%%%%%%%%%%%%% FIG. 1 %%%%%%%%%%%%%%%%%%%%%%%%
\begin{figure}[t]
\centering
\includegraphics[width=8cm]{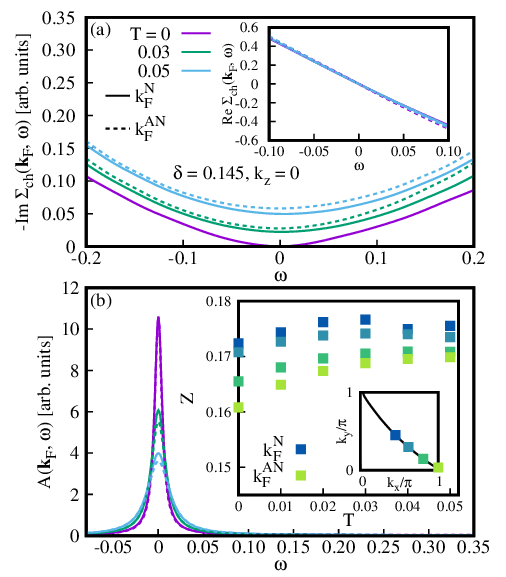}
\caption{Electron property for different temperatures $T$ at the doping rate $\delta=0.145$. (a) Imaginary part of the electron self-energy at the nodal point $\vk_{F}^{N}$ and the antinodal point $\vk_{F}^{AN}$ on the Fermi surface for $T=0.03$ and $0.05$; results at $T=0$ are shown only for $\vk_{F}^{N}$ here. 
The Fermi momenta are defined in panel (b). The inset is the corresponding real part of the self-energy at $T=0$ and $0.05$. (b) Corresponding spectral function for the temperatures in panel (a). The inset describes the quasiparticle weight $Z$ as a function of temperature at several choices of Fermi momenta. 
}
\label{T-depend}
\end{figure}
%%%%%%%%%%%%%%%%%%%%%%%%%%%%%%%%%%%%%%%%%%%%%%%%%

Figure~\ref{T-depend}(a) shows that  Im$\Sigma_{\rm ch}(\vk_{F}, \omega)$---the imaginary part of the electron self-energy from charge fluctuations at the Fermi momentum $\vk_{F}$---vanishes at energy $\omega=0$ and temperature $T=0$ and is characterized by $\sim \omega^{2}$ dependence including the case at finite temperatures; see Appendix~A for a further analysis. In addition, we can check that its temperature dependence at zero energy is characterized by Im$\Sigma_{\rm ch}(\vk_{F}, 0)\sim T^{2}$. In the inset in \fig{T-depend}(a), we plot the corresponding real part Re$\Sigma_{\rm ch}(\vk_{F}, \omega)$. It shows a linear dependence with a negative slope at $\omega=0$, a typical feature of a Fermi liquid. As expected, the spectral function $A(\vk_{F}, \omega)$ exhibits a single peak at $\omega=0$ as shown in \fig{T-depend}(b). All these results demonstrate that despite the presence of acoustic-like plasmon excitations as well as gapless particle-hole excitations, charge fluctuations do not yield a non-Fermi liquid feature, but the system retains the Fermi-liquid property.

However, the quasiparticle weight is reduced substantially. To see this, we compute the quasiparticle weight $Z=(1-\frac{\partial {\rm Re}\Sigma_{\rm ch}(\vk_{F}, \omega)}{\partial \omega} |_{\omega=0})^{-1}$ as a function of temperature in the inset of \fig{T-depend}(b). The value of $Z$ depends weakly on temperature and is around 0.17, meaning that charge fluctuations leave tiny quasiparticle weight around the Fermi energy at all temperatures. It is interesting to note in \fig{T-depend}(b) that the value of $Z$ becomes smaller at lower temperature, but the spectral function becomes sharper at lower temperatures. 

In all panels in \fig{T-depend} (except for Im$\Sigma_{\rm ch}(\vk_{F}, \omega)$ at $T=0$), we plot results for two characteristic momenta,  $\vk_{F}^{N}$ and $\vk_{F}^{AN}$, each of which corresponds to the nodal and antinodal direction [see the inset in \fig{T-depend}(b)]. Although a difference of Im$\Sigma_{\rm ch}(\vk_{F}, \omega)$ between $\vk_{F}^{N}$ and $\vk_{F}^{AN}$ is visible in \fig{T-depend}(a), this is a small effect in the sense that Re$\Sigma_{\rm ch}(\vk_{F}, \omega)$ is not affected practically as seen in the inset in \fig{T-depend}(a). 
In fact, the results in \fig{T-depend}(b) show a weak $\vk_{F}$ dependence. That is, the effect of the charge fluctuations is essentially isotropic, namely $s$-wave-like along the Fermi surface. 

%%%%%%%%%%%%%%%%%%%%% FIG. 2 %%%%%%%%%%%%%%%%%%%%%%%%
\begin{figure}[tb]
\centering
\includegraphics[width=8cm]{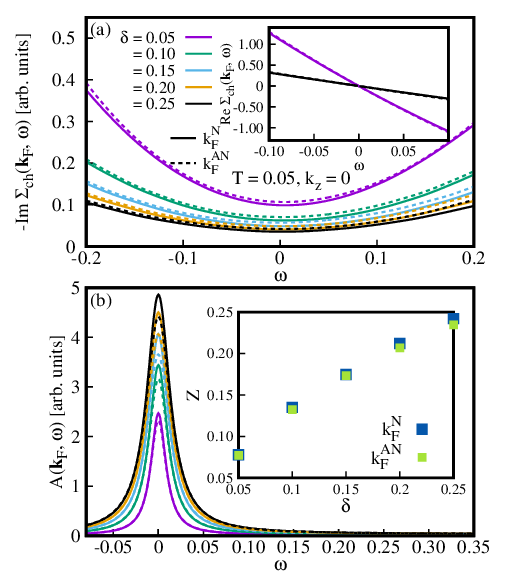}
\caption{Electron property for different doping rates at $T=0.05$. (a) Imaginary part of the self-energy at $\vk_{F}^{N}$ and $\vk_{F}^{AN}$ for $\delta=0.05, 0.10, 0.15, 0.20, 0.25$. The inset shows the corresponding real part of the self-energy at $\delta=0.05$ and 0.25. (b) Spectral function for the doping rates in panel (a). The inset is the doping dependence of $Z$ at $\vk_{F}^{N}$ and $\vk_{F}^{AN}$. 
}
\label{d-depend}
\end{figure}
%%%%%%%%%%%%%%%%%%%%%%%%%%%%%%%%%%%%%%%%%%%%%%%%%

Figure~\ref{d-depend} highlights results of the self-energy for different doping rates at $T=0.05$. In \fig{d-depend}(a), Im$\Sigma_{\rm ch}(\vk_{F}, \omega)$ is characterized by $\sim \omega^{2}$ dependence around $\omega=0$ for all doping rates and the value of Im$\Sigma_{\rm ch}(\vk_{F}, 0)$ decreases with increasing doping. The corresponding results of Re$\Sigma_{\rm ch}(\vk_{F}, \omega)$ are shown in the inset of \fig{d-depend}(a). The slope of Re$\Sigma_{\rm ch}(\vk_{F}, \omega)$ at $\omega=0$ becomes larger with decreasing doping, leading to smaller quasiparticle weight for lower doping---the value of $Z$ varies from $0.08$ to $0.24$ in $0.05 \leq \delta \leq 0.25$ as shown in the inset of \fig{d-depend}(b).  Consequently, the spectral function exhibits a single peak around $\omega=0$ and the peak area becomes smaller with decreasing  doping [\fig{d-depend}(b)]. Interestingly, the peak has a smaller half width at half maximum in spite of a larger value of Im$\Sigma_{\rm ch}(\vk_{F}, \omega)$ with decreasing doping. This counterintuitive behavior is due to a larger negative slope of Re$\Sigma_{\rm ch}(\vk_{F}, \omega)$ around $\omega=0$. It is also intriguing that the self-energy effect from charge fluctuations is pronounced for lower doping in \fig{d-depend}, although the charge degree  of freedom tends to be quenched at half-filling. In all panels in \fig{d-depend}, results do not depend practically on a choice of Fermi momenta.

\subsection{Interplay with the pseudogap}
We have shown that the realistic charge fluctuations in cuprates do not destroy quasiparticles and leave the quasiparticle weight $Z = 0.08$--$0.24$ in $0.05 \leq \delta \leq 0.25$---$Z$ increases with doping. This feature does not depend on temperature nor a choice of Fermi momenta. Given that the present theory captures charge excitation spectra  including plasmons \cite{greco19,greco20,nag20,hepting22}, we expect that the electron self-energy that we have obtained is rather reliable. However, the pseudogap is observed especially in the underdoped region in hole-doped cuprates and the quasiparticle picture is destroyed. What is then a role of charge fluctuations in the presence of the pseudogap? 

We may formally write the self-energy observed in experiments ($\Sigma_{\rm ex}$) as 
\be
\Sigma_{\rm ex} = \Sigma_{\rm ch} + \Sigma_{\rm pg} + \Sigma_{\rm others} \,,
\label{selfex}
\ee
where $\Sigma_{\rm ch}$ is a contribution from the charge fluctuations computed above and $\Sigma_{\rm pg}$ is a component that yields the pseudogap in the spectral function. $\Sigma_{\rm others}$ is the other contributions to the electron self-energy, which may be responsible for strange metallic behavior \cite{mitrano18,husain19,seibold21,caprara22}, a marginal Fermi liquid \cite{varma89}, and other anomalous behavior except for the pseudogap. $\Sigma_{\rm others}$ also contains usual Fermi-liquid corrections from bosonic fluctuations observed in cuprates \cite{carbotte11}. We shall neglect the last term $\Sigma_{\rm others}$ to perform a transparent analysis. 

By  employing a realistic $\Sigma_{\rm ex}$ from experimental data, we may estimate $\Sigma_{\rm pg}$ by modeling it as 
\be
\Sigma_{\rm pg}(\vk, \omega) = \frac{c_{\vk}^{2}}{\omega + i \Gamma_{\vk}} \,. 
\label{selfPG}
\ee
This form is a simplified version capturing consistently various models to describe the pseudogap \cite{norman07}, when focusing on a momentum close to the Fermi surface (see Appendix~B for more details); $\Gamma_{\vk}$ describes a broadening and $c_{\vk}$ has the physical meaning of a kind of gap. Note that as we shall discuss later (see \fig{PG-cond}), the interplay of $c_{\vk}$ and $\Gamma_{\vk}$ is crucial to produce a pseudogap, which has not been recognized much. Since we shall make an analysis by focusing on the antinodal Fermi momentum, we may write $c_{\vk_{F}^{AN}}=c$ and $\Gamma_{\vk_{F}^{AN}}=\Gamma$ for simplicity below.

%%%%%%%%%%%%%%%%%%%%% FIG. 3 %%%%%%%%%%%%%%%%%%%%%%%%
\begin{figure}[tb]
\centering
\includegraphics[width=8cm]{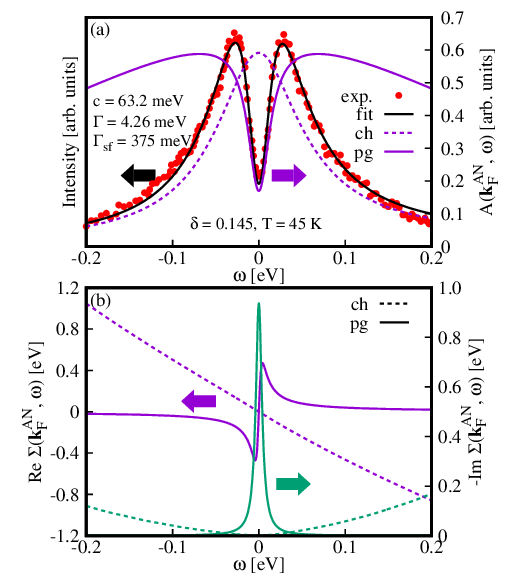}
\caption{Self-energy consistent with experimental data extracted from Ref.~[\onlinecite{kuspert22}]. 
(a) Typical spectral function observed in underdoped cuprates at the antinodal region on the Fermi surface, showing the pseudogap, namely the suppression of the spectral weight at $\omega=0$. The solid black curve is a fitting in terms of Eqs.~(\ref{selfex}) and (\ref{selfPG}); we use $t/2=0.35$ eV \cite{hepting22,misc-factor2a}. 
The spectral function in two different conditions, with only $\Sigma_{\rm ch}$ and with only $\Sigma_{\rm pg}$, is also plotted. 
(b) $\Sigma_{\rm ch}$ and $\Sigma_{\rm pg}$ used in the fitting to the experimental data in (a).
}
\label{PG-fit}
\end{figure}
%%%%%%%%%%%%%%%%%%%%%%%%%%%%%%%%%%%%%%%%%%%%%%%%%

Figure~\ref{PG-fit}(a) is a recent experimental data of the electron spectral function for LSCO \cite{kuspert22} with $\delta=0.145$ at $T=45$ K. We choose the same $\delta$ and $T (=0.0055 t)$ by assuming $t/2=0.35$ eV \cite{hepting22,misc-factor2a}. We then tune the parameters $c$ and $\Gamma$ as well as a broadening of the spectral function $\Gamma_{\rm sf}$ [see \eq{Akw}] to reproduce experimental data [\fig{PG-fit}(a)]. Our obtained $\Sigma_{\rm pg}$ and $\Sigma_{\rm ch}$ are shown in Fig.~\ref{PG-fit}(b). Im$\Sigma_{\rm pg}$ has a sharp peak at $\omega=0$, which generates the pseudogap in \fig{PG-fit}(a). The corresponding Re$\Sigma_{\rm pg}$ exhibits a steep slope with a positive sign at $\omega=0$ to overturn the negative slope of Re$\Sigma_{\rm ch}$. While we have used $\Gamma_{\rm sf}=0.03$ in Figs.~\ref{T-depend} and \ref{d-depend}, we obtain $\Gamma_{\rm sf}=0.536$ to get a better fit especially to the tails away from the Fermi energy in \fig{PG-fit}(a). This large $\Gamma_{\rm sf}$ may also reflect  broadening due to the other contributions $\Sigma_{\rm others}$. 

In \fig{PG-fit}(a) we also plot the spectral function in two different conditions, with only $\Sigma_{\rm pg}$ and with only $\Sigma_{\rm ch}$. While the latter case exhibits a broad, but coherent peak at $\omega=0$---a typical Fermi-liquid feature, the former case indicates that an {\it intrinsic} pseudogap has sizable weight away from $\omega=0$ and forms a very broad structure with a gap nearly double the pseudogap observed in experiments. 

A major surprise in \fig{PG-fit} is that in spite of rather small quasiparticle weight from $\Sigma_{\rm ch}$ ($Z \approx 0.17$ at $\delta=0.145$), we need a very pronounced peak of Im$\Sigma_{\rm pg}$ at $\omega=0$ to reproduce the pseudogap observed experimentally. To explore this outcome more,  we study a condition of $c$ and $\Gamma$ to reproduce a pseudogap in the presence of $\Sigma_{\rm ch}$. We make a map of $\omega_{\rm pg}$---a half distance of double peaks of $A(\vk, \omega)$---in the plane of $c$ and $\Gamma$ in \fig{PG-cond}; $\omega_{\rm pg}=0$ means a single peak at $\omega=0$. The pseudogap is realized below the white curve---this condition is given approximately by (see Appendix~C for an analytical understanding) 
\be
\Gamma^{2} <  2 Z_{\rm FL} c^{2} \,.
\label{PG-eq} 
\ee
Here $Z_{\rm FL}$ is the Fermi-liquid quasiparticle weight in the {\it absence} of $\Sigma_{\rm pg}$ and is given by 0.17 in the present case. The same calculations are also performed for different doping and we superimpose in \fig{PG-cond} the obtained boundary, below which a pseudogap is realized. It shows that we need a severer condition of the choice of $c$ and $\Gamma$ to reproduce the pseudogap for a lower doping rate, where the quasiparticle weight becomes smaller [see \fig{d-depend}(b)]---the contribution $\Sigma_{\rm others}$ in \eq{selfex} that we have neglected would reduce further the value of $Z_{\rm FL}$, yielding a further severer condition of $c$ and $\Gamma$ to produce a pseudogap. From the opposite viewpoint, \fig{PG-cond} indicates that $\Sigma_{\rm pg}$ tends to create a {\it single} peak when the quasiparticle weight becomes {\it smaller} in the absence of $\Sigma_{\rm pg}$. Whether this can be related with the strange metal state in cuprates \cite{keimer15} is an interesting open issue. See Appendix~D for a further analysis. 

%%%%%%%%%%%%%%%%%%%%% FIG. 4 %%%%%%%%%%%%%%%%%%%%%%%%
\begin{figure}[tb]
\centering
\includegraphics[width=8cm]{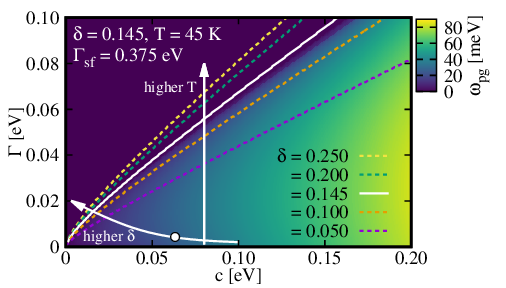}
\caption{Condition to realize the pseudogap in the presence of charge fluctuations in the plane of $c$ and $\Gamma$ in \eq{PG-eq}; we use $t/2=0.35$ eV \cite{hepting22,misc-factor2a}. The pseudogap ($\omega_{\rm pg} \ne 0$) is realized below the white curve.  Similar curves are also superimposed for other doping rates. To be consistent with the pseudogap observed experimentally, $c$ and $\Gamma$ may depend on doping and temperature as shown by arrows schematically. The solid circle corresponds to the values of $c$ and $\Gamma$ used in \fig{PG-fit}. 
}
\label{PG-cond}
\end{figure}
%%%%%%%%%%%%%%%%%%%%%%%%%%%%%%%%%%%%%%%%%%%%%%%%%

\section{Conclusion and discussions}
Recently charge fluctuations were proposed to be responsible for a strange metal and the marginal Fermi-liquid phenomenology \cite{mitrano18,husain19,seibold21,caprara22}. However, we have found that the self-energy from the realistic charge fluctuations is essentially isotropic and yields a Fermi-liquid contribution (Figs.~\ref{T-depend} and \ref{d-depend}). We have also found the small  quasiparticle weight $Z$, which varies from $0.08$ to $0.24$ with increasing doping from $0.05$ to $0.25$ (\fig{d-depend}).  One might expect that a small $Z$ at low doping would work favorably to form a pseudogap because the quasiparticles could be easily destroyed. However, the obtained theoretical insight is the opposite---the smaller the quasiparticle weight is, the more intense additional contributions leading to the pseudogap should be [\eq{PG-eq} and Figs.~\ref{PG-fit} and \ref{PG-cond}]. Furthermore,  to be consistent with experiments, $c$ and $\Gamma$ should exhibit a special doping and temperature dependence as sketched with arrows in \fig{PG-cond}: the gap tends to be closed with decreasing $c$ and to be filled with increasing $\Gamma$---the former feature like a {\it gap-closing} may be caused mainly by increasing doping \cite{damascelli03} and the latter one like a {\it gap-filling} by increasing temperature \cite{norman98a,kanigel07,damascelli03} (see Appendix~E). The microscopic origin of $c$ and $\Gamma$ is a challenge for understanding the pseudogap in cuprates.

In Ref.~\cite{dong19}, a pseudogap very similar to the experimental data was obtained in the dynamical cluster approximation with eight sites to the two-dimensional Hubbard model. However, charge fluctuations in Ref.~[\onlinecite{dong19}] are very different from those reported in RIXS \cite{hepting18,lin20,hepting22,nag20,singh22a} and also very weak. It is  interesting to check whether the reported pseudogap in Ref.~\cite{dong19} practically remains even when the realistic charge fluctuations are taken into account.  

In the overdoped region, we expect $\Sigma_{\rm pg} \rightarrow 0$, but charge fluctuations survive.  The fact that $\Sigma_{\rm ch}$ is essentially isotropic on the Fermi surface (Fig.~\ref{T-depend}) and Im$\Sigma_{\rm ch} \sim T^{2}$ may indicate that $\Sigma_{\rm ch}$ is  promising to describe the transport properties in overdoped cuprates where the scattering rate is isotropic \cite{jawad06,jawad07,french09} and shows a dominant $T^{2}$ dependence \cite{nakamae03,cooper09,harada22,jawad06,jawad07,french09}. In addition, Im$\Sigma_{\rm ch}$ decreases with increasing doping [Fig.~\ref{d-depend}(a)], which is also in line with the behavior of the resistivity \cite{takagi92,timusk99}.

Our value of $Z$ is around 0.25 in the overdoped region [see the inset of \fig{d-depend}(b)], implying the mass enhancement is around 4. Quantum oscillation measurements of ${\rm Tl_2Ba_2CuO_{6+\delta}}$ found a value of 3.1 - 5.1 \cite{vignolle08}, which is consistent with the present work. On the other hand, ARPES measurements for overdoped ${\rm Bi_2Sr_2CaCu_2O_{8+\delta}}$ reported a value around 1.5 \cite{johnson01}. This difference might be related to the difference of the energy scale between quantum oscillation and ARPES. 

The pseudogap in cuprates, namely  $\Sigma_{\rm ex}$, has been frequently studied by focusing on $\Sigma_{\rm pg}$ in \eq{selfex} alone. However, we have demonstrated that the other contributions $\Sigma_{\rm ch}$ and $\Sigma_{\rm others}$ can be crucially important as shown in \fig{PG-fit} and \eq{PG-eq}. Hence it is important to disentangle the source of the pseudogap from $\Sigma_{\rm ex}$. A valuable insight may be obtained as follows. We first approximate $\Sigma_{\rm pg}=0$ around the nodal Fermi momentum, leading to the components of $\Sigma_{\rm ch} + \Sigma_{\rm others}$. Then assuming those components are isotropic, we may obtain $\Sigma_{\rm pg}$ by subtracting the component  $\Sigma_{\rm ch} + \Sigma_{\rm others}$ from $\Sigma_{\rm ex}$ at Fermi momenta away from the nodal point. This procedure may also be performed in numerical calculations as those in Refs.~\cite{gunnarsson15} and \cite{schafer21}.

\acknowledgments
The authors thank L. Manuel, W. Metzner, and T. Sch\"afer for valuable discussions. A part of the results presented in this work was obtained by using the facilities of the CCT-Rosario Computational Center, member of the High Performance Computing National System (SNCAD, MincyT-Argentina). A.G. and H.Y. are indebted to warm hospitality of Max-Planck-Institute for Solid State Research. H.Y. was supported by JSPS KAKENHI Grant No.~JP20H01856 and World Premier International  Research Center Initiative (WPI), MEXT, Japan.

%\bibliographystyle{apsrev4-1}
%\bibliography{main}
%\end{document}

\appendix

\section{\boldmath{$\omega$} dependence of Im\boldmath{$\Sigma$}} 
Here we provide an in-depth analysis of the results in \fig{T-depend}(a) at $T=0$. 

\begin{figure}[t]
\centering
\includegraphics[width=7.5cm]{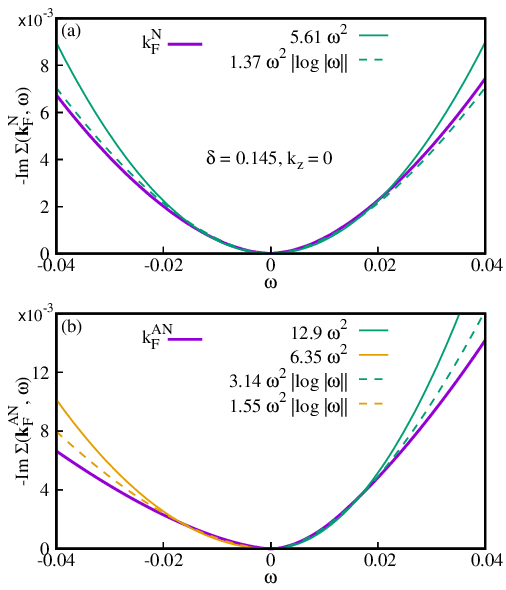}
\caption{$\omega$ dependence of Im$\Sigma(\vk_{F}, \omega)$ at $T=0$. (a) Fitting of Im$\Sigma(\vk_{F}, \omega)$ for $\vk_{F}=\vk_{F}^{N}$ by using two different functional forms, $\omega^{2}$ and $\omega^{2}\log |\omega|$.  (b) Fitting of Im$\Sigma(\vk_{F}, \omega)$ for $\vk_{F}=\vk_{F}^{AN}$ in the positive  and negative energy regions separately. 
}
\label{fit-omega}
\end{figure}

In \fig{fit-omega}(a), our numerical results Im$\Sigma_{\rm ch}(\vk_{F}, \omega)$ at $T=0$  for $\vk_{F}=\vk_{F}^{N}$ are fitted by using two different functional forms, $\omega^{2}$ and $\omega^{2} \log |\omega|$---the former is expected for the three-dimensional (3D) Fermi liquids and the latter for two-dimensional (2D) Fermi liquids \cite{giuliani}. We see both nicely fit to the numerical results in the vicinity of $\omega=0$. Since our system is a layered model, we would expect a crossover from the 2D to the 3D character with decreasing $\omega$ toward zero. Numerically, however, we cannot clearly distinguish them in the vicinity of $\omega=0$. Rather we could firmly say that the 2D character is more pronounced in a higher $\omega$ region. 

%%%%%%%%%%%%%%%%%%%%% FIG.  5 %%%%%%%%%%%%%%%%%%%%%%%%
%%%%%%%%%%%%%%%%%%%%%%%%%%%%%%%%%%%%%%%%%%%%%%%%%

One would expect that Im$\Sigma_{\rm ch}(\vk_{F}, \omega)$ should have a symmetry with respect to $\omega=0$. However, a close inspection in \fig{fit-omega}(a) reveals that this is not exactly the case in the present model. We checked numerically that contributions from the saddle-point regions in $\varepsilon_{\vk - \vq}$ in \eq{ImSig} are larger in the positive $\omega$ side than in the negative $\omega$ side, which we interpret as one of the main sources to yield an asymmetry of Im$\Sigma_{\rm ch}(\vk_{F}, \omega)$ with respect to $\omega=0$ in the low-energy region.

This effect becomes more pronounced when we choose $\vk_{F}=\vk_{F}^{AN}$, much closer to the saddle points for a small $\vq$, as shown in \fig{fit-omega}(b). This was the reason why we refrained from presenting Im$\Sigma_{\rm ch}(\vk_{F}^{AN}, \omega)$ in \fig{T-depend}(a)---special care may be necessary for $\vk_{F}=\vk_{F}^{AN}$ at $T=0$. We thus consider the positive and negative energy regions separately and perform the fitting in each region. We find that the numerical results are well fitted to both $\omega^{2}$ and $\omega^{2}  \log | \omega |$ in the vicinity of $\omega=0$ and the higher energy region is fitted better to the latter.  These technical subtleties, however, are special at $T=0$ especially for $\vk_{F}=\vk_{F}^{AN}$ and fade away at finite temperatures as seen in Figs.~\ref{T-depend}(a) and \ref{d-depend}(a). 

\section{Modeling of the pseudogap} 
The origin of the pseudogap is still controversial and it is beyond the scope of the present work to pursuit it. Instead, from a practical point of view, we consider a self-energy that can reproduce the pseudogap observed by ARPES. 

Our modeling in terms of \eq{selfPG} is based on Ref.~\cite{norman07} and can be regarded as a simplified version to cover different scenarios to capture the pseudogap phenomenology. The self-energy we consider is given by 
\be
\Sigma_{\rm pg}(\vk, \omega) = \frac{c_{\vk}^{2}} {\omega + \tilde\varepsilon_{\vk} + i \Gamma} \,.
\ee
In the case of a commensurate density wave with momentum $\vQ=(\pi, \pi)$ such as the usual charge- and spin-density-wave, $c_{\vk}$ corresponds to its gap and 
\be
\tilde\varepsilon_{\vk}= - \varepsilon_{\vk+ \vQ} \,.
\ee
In the so-called Yang-Rice-Zhang (YRZ) model \cite{yang06}, $c_{\vk}$ controls the magnitude of a pseudogap  and $\tilde\varepsilon_{\vk}$ is the nearest-neighbor term of the tight-binding dispersion 
\be
\tilde\varepsilon_{\vk}= - 2 t (\cos k_{x} + \cos k_{y}) \,. 
\ee
If the $d$-wave pairing fluctuations are responsible for the pseudogap formation, $c_{\vk}$ is the usual $d$-wave pairing gap and we have 
\be
\tilde\varepsilon_{\vk}= \varepsilon_{\vk} \,.
\ee

We consider a momentum fulfilling $\tilde\varepsilon_{\vk}=0$. This condition determines the Luttinger surface, where the self-energy diverges at $\omega=0$ and $\Gamma=+0$. Therefore, the spectral function is expected to be strongly suppressed at zero energy when the Fermi surface crosses the Luttinger surface. In order to capture a pseudogap feature, therefore, we consider a situation where $\tilde\varepsilon_{\vk_{F}} \approx 0$. This is typically realized close to a momentum of the antinodal region in hole-doped cuprates, where a holelike Fermi surface crosses the Brillouin zone boundary. This consideration leads to \eq{selfPG} in the main text after allowing a momentum dependence of $\Gamma$. 

While we have successfully fitted experimental data with \eq{selfPG} (see \fig{PG-fit}), this may not necessarily indicate that the pseudogap should be explained in either of the above three scenarios. This is because the functional form of our simplified self-energy \eq{selfPG} might also be obtained in other scenarios \cite{gunnarsson15,schafer21} and in this sense can be general phenomenologically.  

\section{Analytical understanding of \eq{PG-eq}} 
The condition to produce a pseudogap is given approximately by \eq{PG-eq}. Here we provide analytical grounds behind it. 

Since $\Sigma_{\rm ch}$ exhibits a Fermi-liquid feature in Figs.~\ref{T-depend} and \ref{d-depend}, we may approximate it as
\be
\Sigma_{\rm ch} \approx -a \omega - i \frac{a \pi}{2 \omega_{0}} \omega^{2} \,,
\ee
where $a$ is positive and we have assumed that Im$\Sigma_{\rm ch}=0$ in $| \omega | > \omega_{0}$. This approximation is expected to be good as long as we consider a low-energy property. We then obtain 
\be
Z_{\rm FL}^{-1} =  1-  \frac{\partial {\rm Re} \Sigma_{\rm ch}} {\partial \omega}  =1+a \,.
\ee
Together with $\Sigma_{\rm pg}$ in \eq{selfPG}, we may write the total self-energy as 
\bea
&&\Sigma_{\rm ex} \approx \Sigma_{\rm ch} +  \Sigma_{\rm pg} \\ 
&& \hspace{8mm} \approx -\left( a-\frac{c^{2}}{\omega^{2}+\Gamma^{2}} \right) \omega 
- i \frac{\Gamma c^{2}}{\omega^{2}+\Gamma^{2}} \,.
\eea
Here we have focused on a low-energy region so that Im$\Sigma_{\rm ch}$ is negligible compared with the contribution from Im$\Sigma_{\rm pg}$. We then replace $\Sigma_{\rm ch}$ in \eq{Akw} with the above $\Sigma_{\rm ex}$ and put $\Gamma_{\rm sf}=+0$. At the Fermi momentum $\vk_{F}$, we find that the spectral function $A(\vk_{F}, \omega)$ has a double peak around $\omega=0$ in the condition of 
\be
\Gamma^{2} < 2 Z_{\rm FL} c^{2}\,.
\ee
By comparing with numerical results, we checked that this formula is very precise for $\Gamma_{\rm sf}=0.001$ eV and starts to have visible errors when a larger $\Gamma_{\rm sf}$ is invoked---yet it works as a reliable guide to estimate the boundary of the pseudogap.

\section{Coherent and incoherent single peaks}
%%%%%%%%%%%%%%%%%%%%% FIG. 6 %%%%%%%%%%%%%%%%%%%%%%%%
\begin{figure}[t]
\centering
\includegraphics[width=8cm]{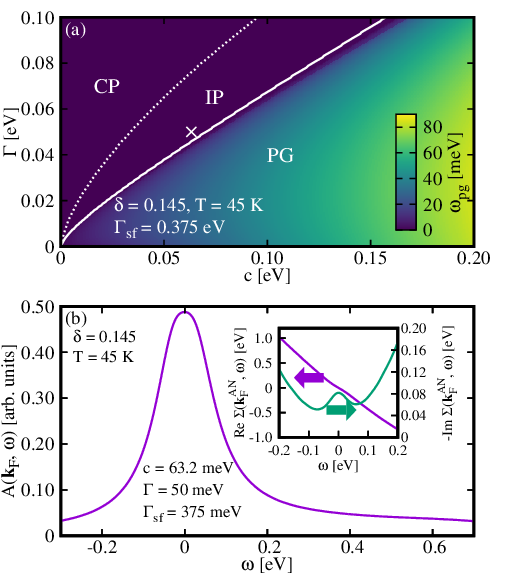}
\caption{Coherent and incoherent single peaks. (a) Reproduction of \fig{PG-cond}, but focusing on the results for $\delta=0.145$. 
The non-pseudogap region above the white curve is divided into two regions, where a coherent or an incoherent single peak is realized. (b) Spectral function at the antinodal Fermi momentum in the IP region marked by the cross in (a). The inset shows that the real part of the self-energy has a negative slope but the imaginary part has a peak at $\omega = 0$.} 
\label{incoherent}
\end{figure}
%%%%%%%%%%%%%%%%%%%%%%%%%%%%%%%%%%%%%%%%%%%%%%%%%

As shown in \fig{PG-cond}, a pseudogap is formed in $\Gamma^{2} \lesssim 2 Z_{\rm FL} c^{2}$ [\eq{PG-eq}]. If this condition is not fulfilled,  a single peak is realized. As indicated in \fig{incoherent}(a), there are two kinds of single peaks. One is a coherent peak (CP) typical of the Fermi liquid as we already discussed in Figs.~\ref{T-depend} and \ref{d-depend}. The other is an incoherent peak (IP) shown in \fig{incoherent}(b), for which Im$\Sigma$ exhibits a peak at $\omega=0$, but Re$\Sigma$ retains a negative slope there as shown in the inset. 

There can be a different IP in that Im$\Sigma$ exhibits a peak at $\omega=0$ and Re$\Sigma$ has a positive slope there, as actually obtained in a theoretical study of electronic nematic fluctuations \cite{yamase12}. 
However, we do not find this kind of IP in \fig{incoherent}(a).  

Figure~\ref{incoherent}(a) indicates that the IP state intervenes between the PG and CP states. Recalling the strange metal state also intervenes between the PG and Fermi-liquid states in cuprates \cite{keimer15}, it is interesting to explore further a possible connection between the IP state and the strange metal state.

\section{Evolution of the spectral function with \boldmath{$c$} and \boldmath{$\Gamma$}} 

%%%%%%%%%%%%%%%%%%%%% FIG. 7 %%%%%%%%%%%%%%%%%%%%%%%%
\begin{figure}[t]
\centering
\includegraphics[width=10cm]{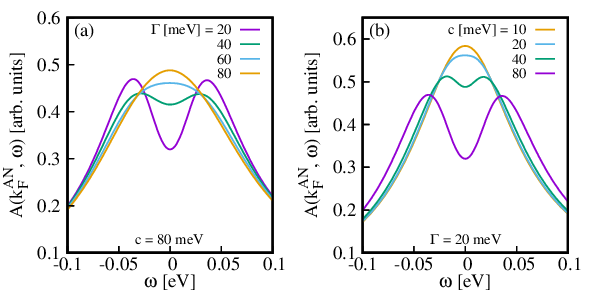}
\caption{Evolution of the spectral function at the antinodal Fermi momentum in the phase diagram shown in \fig{PG-cond}. (a) Several choices of $\Gamma$ at a fixed $c$ and (b) several choices of $c$ at a fixed $\Gamma$. 
}
\label{Akw-Gc}
\end{figure}
%%%%%%%%%%%%%%%%%%%%%%%%%%%%%%%%%%%%%%%%%%%%%%%%%

We have focused on the spectral function fitted to  the experimental data \cite{kuspert22} in the main text. Here from a general point of view, we present how the spectral function evolves by changing $c$ and $\Gamma$ in \fig{PG-cond} at $\delta=0.145$. Figure~\ref{Akw-Gc}(a) shows the spectral function as a function of $\omega$ at the antinodal Fermi momentum. With increasing $\Gamma$ the spectral weight around $\omega=0$ increases while seemingly keeping the gap magnitude. This {\it gap-filling} behavior is typically observed in experiments by increasing temperature \cite{norman98a,kanigel07,damascelli03}. On the other hand, the gap itself is suppressed with decreasing $c$ as shown in \fig{Akw-Gc}(b)---{\it gap-closing} behavior. A similar feature is observed typically when increasing the doping in experiments \cite{damascelli03}. These are underlying considerations to sketch the arrows in \fig{PG-cond}.

\bibliography{main}

\begin{thebibliography}{54}%
\makeatletter
\providecommand \@ifxundefined [1]{%
 \@ifx{#1\undefined}
}%
\providecommand \@ifnum [1]{%
 \ifnum #1\expandafter \@firstoftwo
 \else \expandafter \@secondoftwo
 \fi
}%
\providecommand \@ifx [1]{%
 \ifx #1\expandafter \@firstoftwo
 \else \expandafter \@secondoftwo
 \fi
}%
\providecommand \natexlab [1]{#1}%
\providecommand \enquote  [1]{``#1''}%
\providecommand \bibnamefont  [1]{#1}%
\providecommand \bibfnamefont [1]{#1}%
\providecommand \citenamefont [1]{#1}%
\providecommand \href@noop [0]{\@secondoftwo}%
\providecommand \href [0]{\begingroup \@sanitize@url \@href}%
\providecommand \@href[1]{\@@startlink{#1}\@@href}%
\providecommand \@@href[1]{\endgroup#1\@@endlink}%
\providecommand \@sanitize@url [0]{\catcode `\\12\catcode `\$12\catcode
  `\&12\catcode `\#12\catcode `\^12\catcode `\_12\catcode `\%12\relax}%
\providecommand \@@startlink[1]{}%
\providecommand \@@endlink[0]{}%
\providecommand \url  [0]{\begingroup\@sanitize@url \@url }%
\providecommand \@url [1]{\endgroup\@href {#1}{\urlprefix }}%
\providecommand \urlprefix  [0]{URL }%
\providecommand \Eprint [0]{\href }%
\providecommand \doibase [0]{http://dx.doi.org/}%
\providecommand \selectlanguage [0]{\@gobble}%
\providecommand \bibinfo  [0]{\@secondoftwo}%
\providecommand \bibfield  [0]{\@secondoftwo}%
\providecommand \translation [1]{[#1]}%
\providecommand \BibitemOpen [0]{}%
\providecommand \bibitemStop [0]{}%
\providecommand \bibitemNoStop [0]{.\EOS\space}%
\providecommand \EOS [0]{\spacefactor3000\relax}%
\providecommand \BibitemShut  [1]{\csname bibitem#1\endcsname}%
\let\auto@bib@innerbib\@empty
%</preamble>
\bibitem [{\citenamefont {Damascelli}\ \emph {et~al.}(2003)\citenamefont
  {Damascelli}, \citenamefont {Hussain},\ and\ \citenamefont
  {Shen}}]{damascelli03}%
  \BibitemOpen
  \bibfield  {author} {\bibinfo {author} {\bibfnamefont {A.}~\bibnamefont
  {Damascelli}}, \bibinfo {author} {\bibfnamefont {Z.}~\bibnamefont {Hussain}},
  \ and\ \bibinfo {author} {\bibfnamefont {Z.-X.}\ \bibnamefont {Shen}},\
  }\href {\doibase 10.1103/RevModPhys.75.473} {\bibfield  {journal} {\bibinfo
  {journal} {Rev. Mod. Phys.}\ }\textbf {\bibinfo {volume} {75}},\ \bibinfo
  {pages} {473} (\bibinfo {year} {2003})}\BibitemShut {NoStop}%
\bibitem [{\citenamefont {Timusk}\ and\ \citenamefont
  {Statt}(1999)}]{timusk99}%
  \BibitemOpen
  \bibfield  {author} {\bibinfo {author} {\bibfnamefont {T.}~\bibnamefont
  {Timusk}}\ and\ \bibinfo {author} {\bibfnamefont {B.}~\bibnamefont {Statt}},\
  }\href {\doibase 10.1088/0034-4885/62/1/002} {\bibfield  {journal} {\bibinfo
  {journal} {Reports on Progress in Physics}\ }\textbf {\bibinfo {volume}
  {62}},\ \bibinfo {pages} {61} (\bibinfo {year} {1999})}\BibitemShut {NoStop}%
\bibitem [{\citenamefont {Keimer}\ \emph {et~al.}(2015)\citenamefont {Keimer},
  \citenamefont {Kivelson}, \citenamefont {Norman}, \citenamefont {Uchida},\
  and\ \citenamefont {Zaanen}}]{keimer15}%
  \BibitemOpen
  \bibfield  {author} {\bibinfo {author} {\bibfnamefont {B.}~\bibnamefont
  {Keimer}}, \bibinfo {author} {\bibfnamefont {S.~A.}\ \bibnamefont
  {Kivelson}}, \bibinfo {author} {\bibfnamefont {M.~R.}\ \bibnamefont
  {Norman}}, \bibinfo {author} {\bibfnamefont {S.}~\bibnamefont {Uchida}}, \
  and\ \bibinfo {author} {\bibfnamefont {J.}~\bibnamefont {Zaanen}},\ }\href
  {\doibase 10.1038/nature14165} {\bibfield  {journal} {\bibinfo  {journal}
  {Nature}\ }\textbf {\bibinfo {volume} {518}},\ \bibinfo {pages} {179}
  (\bibinfo {year} {2015})}\BibitemShut {NoStop}%
\bibitem [{\citenamefont {Hepting}\ \emph {et~al.}(2018)\citenamefont
  {Hepting}, \citenamefont {Chaix}, \citenamefont {Huang}, \citenamefont
  {Fumagalli}, \citenamefont {Peng}, \citenamefont {Moritz}, \citenamefont
  {Kummer}, \citenamefont {Brookes}, \citenamefont {Lee}, \citenamefont
  {Hashimoto}, \citenamefont {Sarkar}, \citenamefont {He}, \citenamefont
  {Rotundu}, \citenamefont {Lee}, \citenamefont {Greene}, \citenamefont
  {Braicovich}, \citenamefont {Ghiringhelli}, \citenamefont {Shen},
  \citenamefont {Devereaux},\ and\ \citenamefont {Lee}}]{hepting18}%
  \BibitemOpen
  \bibfield  {author} {\bibinfo {author} {\bibfnamefont {M.}~\bibnamefont
  {Hepting}}, \bibinfo {author} {\bibfnamefont {L.}~\bibnamefont {Chaix}},
  \bibinfo {author} {\bibfnamefont {E.~W.}\ \bibnamefont {Huang}}, \bibinfo
  {author} {\bibfnamefont {R.}~\bibnamefont {Fumagalli}}, \bibinfo {author}
  {\bibfnamefont {Y.~Y.}\ \bibnamefont {Peng}}, \bibinfo {author}
  {\bibfnamefont {B.}~\bibnamefont {Moritz}}, \bibinfo {author} {\bibfnamefont
  {K.}~\bibnamefont {Kummer}}, \bibinfo {author} {\bibfnamefont {N.~B.}\
  \bibnamefont {Brookes}}, \bibinfo {author} {\bibfnamefont {W.~C.}\
  \bibnamefont {Lee}}, \bibinfo {author} {\bibfnamefont {M.}~\bibnamefont
  {Hashimoto}}, \bibinfo {author} {\bibfnamefont {T.}~\bibnamefont {Sarkar}},
  \bibinfo {author} {\bibfnamefont {J.-F.}\ \bibnamefont {He}}, \bibinfo
  {author} {\bibfnamefont {C.~R.}\ \bibnamefont {Rotundu}}, \bibinfo {author}
  {\bibfnamefont {Y.~S.}\ \bibnamefont {Lee}}, \bibinfo {author} {\bibfnamefont
  {R.~L.}\ \bibnamefont {Greene}}, \bibinfo {author} {\bibfnamefont
  {L.}~\bibnamefont {Braicovich}}, \bibinfo {author} {\bibfnamefont
  {G.}~\bibnamefont {Ghiringhelli}}, \bibinfo {author} {\bibfnamefont {Z.~X.}\
  \bibnamefont {Shen}}, \bibinfo {author} {\bibfnamefont {T.~P.}\ \bibnamefont
  {Devereaux}}, \ and\ \bibinfo {author} {\bibfnamefont {W.~S.}\ \bibnamefont
  {Lee}},\ }\href {\doibase 10.1038/s41586-018-0648-3} {\bibfield  {journal}
  {\bibinfo  {journal} {Nature}\ }\textbf {\bibinfo {volume} {563}},\ \bibinfo
  {pages} {374} (\bibinfo {year} {2018})}\BibitemShut {NoStop}%
\bibitem [{\citenamefont {Lin}\ \emph {et~al.}(2020)\citenamefont {Lin},
  \citenamefont {Yuan}, \citenamefont {Jin}, \citenamefont {Yin}, \citenamefont
  {Li}, \citenamefont {Zhou}, \citenamefont {Lu}, \citenamefont {Dantz},
  \citenamefont {Schmitt}, \citenamefont {Ding}, \citenamefont {Guo},
  \citenamefont {Dean},\ and\ \citenamefont {Liu}}]{lin20}%
  \BibitemOpen
  \bibfield  {author} {\bibinfo {author} {\bibfnamefont {J.}~\bibnamefont
  {Lin}}, \bibinfo {author} {\bibfnamefont {J.}~\bibnamefont {Yuan}}, \bibinfo
  {author} {\bibfnamefont {K.}~\bibnamefont {Jin}}, \bibinfo {author}
  {\bibfnamefont {Z.}~\bibnamefont {Yin}}, \bibinfo {author} {\bibfnamefont
  {G.}~\bibnamefont {Li}}, \bibinfo {author} {\bibfnamefont {K.-J.}\
  \bibnamefont {Zhou}}, \bibinfo {author} {\bibfnamefont {X.}~\bibnamefont
  {Lu}}, \bibinfo {author} {\bibfnamefont {M.}~\bibnamefont {Dantz}}, \bibinfo
  {author} {\bibfnamefont {T.}~\bibnamefont {Schmitt}}, \bibinfo {author}
  {\bibfnamefont {H.}~\bibnamefont {Ding}}, \bibinfo {author} {\bibfnamefont
  {H.}~\bibnamefont {Guo}}, \bibinfo {author} {\bibfnamefont {M.~P.~M.}\
  \bibnamefont {Dean}}, \ and\ \bibinfo {author} {\bibfnamefont
  {X.}~\bibnamefont {Liu}},\ }\href {\doibase 10.1038/s41535-019-0205-9}
  {\bibfield  {journal} {\bibinfo  {journal} {npj Quantum Materials}\ }\textbf
  {\bibinfo {volume} {5}},\ \bibinfo {pages} {4} (\bibinfo {year}
  {2020})}\BibitemShut {NoStop}%
\bibitem [{\citenamefont {Hepting}\ \emph {et~al.}(2022)\citenamefont
  {Hepting}, \citenamefont {Bejas}, \citenamefont {Nag}, \citenamefont
  {Yamase}, \citenamefont {Coppola}, \citenamefont {Betto}, \citenamefont
  {Falter}, \citenamefont {Garcia-Fernandez}, \citenamefont {Agrestini},
  \citenamefont {Zhou}, \citenamefont {Minola}, \citenamefont {Sacco},
  \citenamefont {Maritato}, \citenamefont {Orgiani}, \citenamefont {Wei},
  \citenamefont {Shen}, \citenamefont {Schlom}, \citenamefont {Galdi},
  \citenamefont {Greco},\ and\ \citenamefont {Keimer}}]{hepting22}%
  \BibitemOpen
  \bibfield  {author} {\bibinfo {author} {\bibfnamefont {M.}~\bibnamefont
  {Hepting}}, \bibinfo {author} {\bibfnamefont {M.}~\bibnamefont {Bejas}},
  \bibinfo {author} {\bibfnamefont {A.}~\bibnamefont {Nag}}, \bibinfo {author}
  {\bibfnamefont {H.}~\bibnamefont {Yamase}}, \bibinfo {author} {\bibfnamefont
  {N.}~\bibnamefont {Coppola}}, \bibinfo {author} {\bibfnamefont
  {D.}~\bibnamefont {Betto}}, \bibinfo {author} {\bibfnamefont
  {C.}~\bibnamefont {Falter}}, \bibinfo {author} {\bibfnamefont
  {M.}~\bibnamefont {Garcia-Fernandez}}, \bibinfo {author} {\bibfnamefont
  {S.}~\bibnamefont {Agrestini}}, \bibinfo {author} {\bibfnamefont {K.-J.}\
  \bibnamefont {Zhou}}, \bibinfo {author} {\bibfnamefont {M.}~\bibnamefont
  {Minola}}, \bibinfo {author} {\bibfnamefont {C.}~\bibnamefont {Sacco}},
  \bibinfo {author} {\bibfnamefont {L.}~\bibnamefont {Maritato}}, \bibinfo
  {author} {\bibfnamefont {P.}~\bibnamefont {Orgiani}}, \bibinfo {author}
  {\bibfnamefont {H.~I.}\ \bibnamefont {Wei}}, \bibinfo {author} {\bibfnamefont
  {K.~M.}\ \bibnamefont {Shen}}, \bibinfo {author} {\bibfnamefont {D.~G.}\
  \bibnamefont {Schlom}}, \bibinfo {author} {\bibfnamefont {A.}~\bibnamefont
  {Galdi}}, \bibinfo {author} {\bibfnamefont {A.}~\bibnamefont {Greco}}, \ and\
  \bibinfo {author} {\bibfnamefont {B.}~\bibnamefont {Keimer}},\ }\href
  {\doibase 10.1103/PhysRevLett.129.047001} {\bibfield  {journal} {\bibinfo
  {journal} {Phys. Rev. Lett.}\ }\textbf {\bibinfo {volume} {129}},\ \bibinfo
  {pages} {047001} (\bibinfo {year} {2022})}\BibitemShut {NoStop}%
\bibitem [{\citenamefont {Nag}\ \emph {et~al.}(2020)\citenamefont {Nag},
  \citenamefont {Zhu}, \citenamefont {Bejas}, \citenamefont {Li}, \citenamefont
  {Robarts}, \citenamefont {Yamase}, \citenamefont {Petsch}, \citenamefont
  {Song}, \citenamefont {Eisaki}, \citenamefont {Walters}, \citenamefont
  {Garc\'{\i}a-Fern\'andez}, \citenamefont {Greco}, \citenamefont {Hayden},\
  and\ \citenamefont {Zhou}}]{nag20}%
  \BibitemOpen
  \bibfield  {author} {\bibinfo {author} {\bibfnamefont {A.}~\bibnamefont
  {Nag}}, \bibinfo {author} {\bibfnamefont {M.}~\bibnamefont {Zhu}}, \bibinfo
  {author} {\bibfnamefont {M.}~\bibnamefont {Bejas}}, \bibinfo {author}
  {\bibfnamefont {J.}~\bibnamefont {Li}}, \bibinfo {author} {\bibfnamefont
  {H.~C.}\ \bibnamefont {Robarts}}, \bibinfo {author} {\bibfnamefont
  {H.}~\bibnamefont {Yamase}}, \bibinfo {author} {\bibfnamefont {A.~N.}\
  \bibnamefont {Petsch}}, \bibinfo {author} {\bibfnamefont {D.}~\bibnamefont
  {Song}}, \bibinfo {author} {\bibfnamefont {H.}~\bibnamefont {Eisaki}},
  \bibinfo {author} {\bibfnamefont {A.~C.}\ \bibnamefont {Walters}}, \bibinfo
  {author} {\bibfnamefont {M.}~\bibnamefont {Garc\'{\i}a-Fern\'andez}},
  \bibinfo {author} {\bibfnamefont {A.}~\bibnamefont {Greco}}, \bibinfo
  {author} {\bibfnamefont {S.~M.}\ \bibnamefont {Hayden}}, \ and\ \bibinfo
  {author} {\bibfnamefont {K.-J.}\ \bibnamefont {Zhou}},\ }\href {\doibase
  10.1103/PhysRevLett.125.257002} {\bibfield  {journal} {\bibinfo  {journal}
  {Phys. Rev. Lett.}\ }\textbf {\bibinfo {volume} {125}},\ \bibinfo {pages}
  {257002} (\bibinfo {year} {2020})}\BibitemShut {NoStop}%
\bibitem [{\citenamefont {Singh}\ \emph {et~al.}(2022)\citenamefont {Singh},
  \citenamefont {Huang}, \citenamefont {Lane}, \citenamefont {Li},
  \citenamefont {Okamoto}, \citenamefont {Komiya}, \citenamefont {Markiewicz},
  \citenamefont {Bansil}, \citenamefont {Lee}, \citenamefont {Fujimori},
  \citenamefont {Chen},\ and\ \citenamefont {Huang}}]{singh22a}%
  \BibitemOpen
  \bibfield  {author} {\bibinfo {author} {\bibfnamefont {A.}~\bibnamefont
  {Singh}}, \bibinfo {author} {\bibfnamefont {H.~Y.}\ \bibnamefont {Huang}},
  \bibinfo {author} {\bibfnamefont {C.}~\bibnamefont {Lane}}, \bibinfo {author}
  {\bibfnamefont {J.~H.}\ \bibnamefont {Li}}, \bibinfo {author} {\bibfnamefont
  {J.}~\bibnamefont {Okamoto}}, \bibinfo {author} {\bibfnamefont
  {S.}~\bibnamefont {Komiya}}, \bibinfo {author} {\bibfnamefont {R.~S.}\
  \bibnamefont {Markiewicz}}, \bibinfo {author} {\bibfnamefont
  {A.}~\bibnamefont {Bansil}}, \bibinfo {author} {\bibfnamefont {T.~K.}\
  \bibnamefont {Lee}}, \bibinfo {author} {\bibfnamefont {A.}~\bibnamefont
  {Fujimori}}, \bibinfo {author} {\bibfnamefont {C.~T.}\ \bibnamefont {Chen}},
  \ and\ \bibinfo {author} {\bibfnamefont {D.~J.}\ \bibnamefont {Huang}},\
  }\href {\doibase 10.1103/PhysRevB.105.235105} {\bibfield  {journal} {\bibinfo
   {journal} {Phys. Rev. B}\ }\textbf {\bibinfo {volume} {105}},\ \bibinfo
  {pages} {235105} (\bibinfo {year} {2022})}\BibitemShut {NoStop}%
\bibitem [{\citenamefont {Greco}\ \emph {et~al.}(2016)\citenamefont {Greco},
  \citenamefont {Yamase},\ and\ \citenamefont {Bejas}}]{greco16}%
  \BibitemOpen
  \bibfield  {author} {\bibinfo {author} {\bibfnamefont {A.}~\bibnamefont
  {Greco}}, \bibinfo {author} {\bibfnamefont {H.}~\bibnamefont {Yamase}}, \
  and\ \bibinfo {author} {\bibfnamefont {M.}~\bibnamefont {Bejas}},\ }\href
  {\doibase 10.1103/PhysRevB.94.075139} {\bibfield  {journal} {\bibinfo
  {journal} {Phys. Rev. B}\ }\textbf {\bibinfo {volume} {94}},\ \bibinfo
  {pages} {075139} (\bibinfo {year} {2016})}\BibitemShut {NoStop}%
\bibitem [{\citenamefont {Dong}\ \emph {et~al.}(2019)\citenamefont {Dong},
  \citenamefont {Chen},\ and\ \citenamefont {Gull}}]{dong19}%
  \BibitemOpen
  \bibfield  {author} {\bibinfo {author} {\bibfnamefont {X.}~\bibnamefont
  {Dong}}, \bibinfo {author} {\bibfnamefont {X.}~\bibnamefont {Chen}}, \ and\
  \bibinfo {author} {\bibfnamefont {E.}~\bibnamefont {Gull}},\ }\href {\doibase
  10.1103/PhysRevB.100.235107} {\bibfield  {journal} {\bibinfo  {journal}
  {Phys. Rev. B}\ }\textbf {\bibinfo {volume} {100}},\ \bibinfo {pages}
  {235107} (\bibinfo {year} {2019})}\BibitemShut {NoStop}%
\bibitem [{\citenamefont {Gunnarsson}\ \emph {et~al.}(2015)\citenamefont
  {Gunnarsson}, \citenamefont {Sch\"afer}, \citenamefont {LeBlanc},
  \citenamefont {Gull}, \citenamefont {Merino}, \citenamefont {Sangiovanni},
  \citenamefont {Rohringer},\ and\ \citenamefont {Toschi}}]{gunnarsson15}%
  \BibitemOpen
  \bibfield  {author} {\bibinfo {author} {\bibfnamefont {O.}~\bibnamefont
  {Gunnarsson}}, \bibinfo {author} {\bibfnamefont {T.}~\bibnamefont
  {Sch\"afer}}, \bibinfo {author} {\bibfnamefont {J.~P.~F.}\ \bibnamefont
  {LeBlanc}}, \bibinfo {author} {\bibfnamefont {E.}~\bibnamefont {Gull}},
  \bibinfo {author} {\bibfnamefont {J.}~\bibnamefont {Merino}}, \bibinfo
  {author} {\bibfnamefont {G.}~\bibnamefont {Sangiovanni}}, \bibinfo {author}
  {\bibfnamefont {G.}~\bibnamefont {Rohringer}}, \ and\ \bibinfo {author}
  {\bibfnamefont {A.}~\bibnamefont {Toschi}},\ }\href {\doibase
  10.1103/PhysRevLett.114.236402} {\bibfield  {journal} {\bibinfo  {journal}
  {Phys. Rev. Lett.}\ }\textbf {\bibinfo {volume} {114}},\ \bibinfo {pages}
  {236402} (\bibinfo {year} {2015})}\BibitemShut {NoStop}%
\bibitem [{\citenamefont {Sch\"afer}\ \emph {et~al.}(2021)\citenamefont
  {Sch\"afer}, \citenamefont {Wentzell}, \citenamefont {\ifmmode~\check{S}\else
  \v{S}\fi{}imkovic}, \citenamefont {He}, \citenamefont {Hille}, \citenamefont
  {Klett}, \citenamefont {Eckhardt}, \citenamefont {Arzhang}, \citenamefont
  {Harkov}, \citenamefont {Le~R\'egent}, \citenamefont {Kirsch}, \citenamefont
  {Wang}, \citenamefont {Kim}, \citenamefont {Kozik}, \citenamefont {Stepanov},
  \citenamefont {Kauch}, \citenamefont {Andergassen}, \citenamefont {Hansmann},
  \citenamefont {Rohe}, \citenamefont {Vilk}, \citenamefont {LeBlanc},
  \citenamefont {Zhang}, \citenamefont {Tremblay}, \citenamefont {Ferrero},
  \citenamefont {Parcollet},\ and\ \citenamefont {Georges}}]{schafer21}%
  \BibitemOpen
  \bibfield  {author} {\bibinfo {author} {\bibfnamefont {T.}~\bibnamefont
  {Sch\"afer}}, \bibinfo {author} {\bibfnamefont {N.}~\bibnamefont {Wentzell}},
  \bibinfo {author} {\bibfnamefont {F.}~\bibnamefont {\ifmmode~\check{S}\else
  \v{S}\fi{}imkovic}}, \bibinfo {author} {\bibfnamefont {Y.-Y.}\ \bibnamefont
  {He}}, \bibinfo {author} {\bibfnamefont {C.}~\bibnamefont {Hille}}, \bibinfo
  {author} {\bibfnamefont {M.}~\bibnamefont {Klett}}, \bibinfo {author}
  {\bibfnamefont {C.~J.}\ \bibnamefont {Eckhardt}}, \bibinfo {author}
  {\bibfnamefont {B.}~\bibnamefont {Arzhang}}, \bibinfo {author} {\bibfnamefont
  {V.}~\bibnamefont {Harkov}}, \bibinfo {author} {\bibfnamefont {F.~m. c.-M.}\
  \bibnamefont {Le~R\'egent}}, \bibinfo {author} {\bibfnamefont
  {A.}~\bibnamefont {Kirsch}}, \bibinfo {author} {\bibfnamefont
  {Y.}~\bibnamefont {Wang}}, \bibinfo {author} {\bibfnamefont {A.~J.}\
  \bibnamefont {Kim}}, \bibinfo {author} {\bibfnamefont {E.}~\bibnamefont
  {Kozik}}, \bibinfo {author} {\bibfnamefont {E.~A.}\ \bibnamefont {Stepanov}},
  \bibinfo {author} {\bibfnamefont {A.}~\bibnamefont {Kauch}}, \bibinfo
  {author} {\bibfnamefont {S.}~\bibnamefont {Andergassen}}, \bibinfo {author}
  {\bibfnamefont {P.}~\bibnamefont {Hansmann}}, \bibinfo {author}
  {\bibfnamefont {D.}~\bibnamefont {Rohe}}, \bibinfo {author} {\bibfnamefont
  {Y.~M.}\ \bibnamefont {Vilk}}, \bibinfo {author} {\bibfnamefont {J.~P.~F.}\
  \bibnamefont {LeBlanc}}, \bibinfo {author} {\bibfnamefont {S.}~\bibnamefont
  {Zhang}}, \bibinfo {author} {\bibfnamefont {A.-M.~S.}\ \bibnamefont
  {Tremblay}}, \bibinfo {author} {\bibfnamefont {M.}~\bibnamefont {Ferrero}},
  \bibinfo {author} {\bibfnamefont {O.}~\bibnamefont {Parcollet}}, \ and\
  \bibinfo {author} {\bibfnamefont {A.}~\bibnamefont {Georges}},\ }\href
  {\doibase 10.1103/PhysRevX.11.011058} {\bibfield  {journal} {\bibinfo
  {journal} {Phys. Rev. X}\ }\textbf {\bibinfo {volume} {11}},\ \bibinfo
  {pages} {011058} (\bibinfo {year} {2021})}\BibitemShut {NoStop}%
\bibitem [{\citenamefont {Yu}\ \emph {et~al.}(2024)\citenamefont {Yu},
  \citenamefont {Iskakov}, \citenamefont {Gull}, \citenamefont {Held},\ and\
  \citenamefont {Krien}}]{yu24}%
  \BibitemOpen
  \bibfield  {author} {\bibinfo {author} {\bibfnamefont {Y.}~\bibnamefont
  {Yu}}, \bibinfo {author} {\bibfnamefont {S.}~\bibnamefont {Iskakov}},
  \bibinfo {author} {\bibfnamefont {E.}~\bibnamefont {Gull}}, \bibinfo {author}
  {\bibfnamefont {K.}~\bibnamefont {Held}}, \ and\ \bibinfo {author}
  {\bibfnamefont {F.}~\bibnamefont {Krien}},\ }\href@noop {} {\enquote
  {\bibinfo {title} {{Unambiguous fluctuation decomposition of the self-energy:
  pseudogap physics beyond spin fluctuations}},}\ } (\bibinfo {year} {2024}),\
  \Eprint {http://arxiv.org/abs/2401.08543} {arXiv:2401.08543
  [cond-mat.str-el]} \BibitemShut {NoStop}%
\bibitem [{\citenamefont {Patel}\ and\ \citenamefont
  {Sachdev}(2019)}]{patel19}%
  \BibitemOpen
  \bibfield  {author} {\bibinfo {author} {\bibfnamefont {A.~A.}\ \bibnamefont
  {Patel}}\ and\ \bibinfo {author} {\bibfnamefont {S.}~\bibnamefont
  {Sachdev}},\ }\href {\doibase 10.1103/PhysRevLett.123.066601} {\bibfield
  {journal} {\bibinfo  {journal} {Phys. Rev. Lett.}\ }\textbf {\bibinfo
  {volume} {123}},\ \bibinfo {pages} {066601} (\bibinfo {year}
  {2019})}\BibitemShut {NoStop}%
\bibitem [{\citenamefont {Grissonnanche}\ \emph {et~al.}(2021)\citenamefont
  {Grissonnanche}, \citenamefont {Fang}, \citenamefont {Legros}, \citenamefont
  {Verret}, \citenamefont {Lalibert{\'e}}, \citenamefont {Collignon},
  \citenamefont {Zhou}, \citenamefont {Graf}, \citenamefont {Goddard},
  \citenamefont {Taillefer},\ and\ \citenamefont {Ramshaw}}]{grissonnanche21}%
  \BibitemOpen
  \bibfield  {author} {\bibinfo {author} {\bibfnamefont {G.}~\bibnamefont
  {Grissonnanche}}, \bibinfo {author} {\bibfnamefont {Y.}~\bibnamefont {Fang}},
  \bibinfo {author} {\bibfnamefont {A.}~\bibnamefont {Legros}}, \bibinfo
  {author} {\bibfnamefont {S.}~\bibnamefont {Verret}}, \bibinfo {author}
  {\bibfnamefont {F.}~\bibnamefont {Lalibert{\'e}}}, \bibinfo {author}
  {\bibfnamefont {C.}~\bibnamefont {Collignon}}, \bibinfo {author}
  {\bibfnamefont {J.}~\bibnamefont {Zhou}}, \bibinfo {author} {\bibfnamefont
  {D.}~\bibnamefont {Graf}}, \bibinfo {author} {\bibfnamefont {P.~A.}\
  \bibnamefont {Goddard}}, \bibinfo {author} {\bibfnamefont {L.}~\bibnamefont
  {Taillefer}}, \ and\ \bibinfo {author} {\bibfnamefont {B.~J.}\ \bibnamefont
  {Ramshaw}},\ }\href {\doibase 10.1038/s41586-021-03697-8} {\bibfield
  {journal} {\bibinfo  {journal} {Nature}\ }\textbf {\bibinfo {volume} {595}},\
  \bibinfo {pages} {667} (\bibinfo {year} {2021})}\BibitemShut {NoStop}%
\bibitem [{\citenamefont {Phillips}\ \emph {et~al.}(2022)\citenamefont
  {Phillips}, \citenamefont {Hussey},\ and\ \citenamefont
  {Abbamonte}}]{phillips22}%
  \BibitemOpen
  \bibfield  {author} {\bibinfo {author} {\bibfnamefont {P.~W.}\ \bibnamefont
  {Phillips}}, \bibinfo {author} {\bibfnamefont {N.~E.}\ \bibnamefont
  {Hussey}}, \ and\ \bibinfo {author} {\bibfnamefont {P.}~\bibnamefont
  {Abbamonte}},\ }\href {\doibase 10.1126/science.abh4273} {\bibfield
  {journal} {\bibinfo  {journal} {Science}\ }\textbf {\bibinfo {volume}
  {377}},\ \bibinfo {pages} {eabh4273} (\bibinfo {year} {2022})}\BibitemShut
  {NoStop}%
\bibitem [{\citenamefont {Mitrano}\ \emph {et~al.}(2018)\citenamefont
  {Mitrano}, \citenamefont {Husain}, \citenamefont {Vig}, \citenamefont
  {Kogar}, \citenamefont {Rak}, \citenamefont {Rubeck}, \citenamefont
  {Schmalian}, \citenamefont {Uchoa}, \citenamefont {Schneeloch}, \citenamefont
  {Zhong}, \citenamefont {Gu},\ and\ \citenamefont {Abbamonte}}]{mitrano18}%
  \BibitemOpen
  \bibfield  {author} {\bibinfo {author} {\bibfnamefont {M.}~\bibnamefont
  {Mitrano}}, \bibinfo {author} {\bibfnamefont {A.~A.}\ \bibnamefont {Husain}},
  \bibinfo {author} {\bibfnamefont {S.}~\bibnamefont {Vig}}, \bibinfo {author}
  {\bibfnamefont {A.}~\bibnamefont {Kogar}}, \bibinfo {author} {\bibfnamefont
  {M.~S.}\ \bibnamefont {Rak}}, \bibinfo {author} {\bibfnamefont {S.~I.}\
  \bibnamefont {Rubeck}}, \bibinfo {author} {\bibfnamefont {J.}~\bibnamefont
  {Schmalian}}, \bibinfo {author} {\bibfnamefont {B.}~\bibnamefont {Uchoa}},
  \bibinfo {author} {\bibfnamefont {J.}~\bibnamefont {Schneeloch}}, \bibinfo
  {author} {\bibfnamefont {R.}~\bibnamefont {Zhong}}, \bibinfo {author}
  {\bibfnamefont {G.~D.}\ \bibnamefont {Gu}}, \ and\ \bibinfo {author}
  {\bibfnamefont {P.}~\bibnamefont {Abbamonte}},\ }\href {\doibase
  10.1073/pnas.1721495115} {\bibfield  {journal} {\bibinfo  {journal} {Proc.
  Natl. Acad. Sci. U. S. A.}\ }\textbf {\bibinfo {volume} {115}},\ \bibinfo
  {pages} {5392} (\bibinfo {year} {2018})}\BibitemShut {NoStop}%
\bibitem [{\citenamefont {Husain}\ \emph {et~al.}(2019)\citenamefont {Husain},
  \citenamefont {Mitrano}, \citenamefont {Rak}, \citenamefont {Rubeck},
  \citenamefont {Uchoa}, \citenamefont {March}, \citenamefont {Dwyer},
  \citenamefont {Schneeloch}, \citenamefont {Zhong}, \citenamefont {Gu},\ and\
  \citenamefont {Abbamonte}}]{husain19}%
  \BibitemOpen
  \bibfield  {author} {\bibinfo {author} {\bibfnamefont {A.~A.}\ \bibnamefont
  {Husain}}, \bibinfo {author} {\bibfnamefont {M.}~\bibnamefont {Mitrano}},
  \bibinfo {author} {\bibfnamefont {M.~S.}\ \bibnamefont {Rak}}, \bibinfo
  {author} {\bibfnamefont {S.}~\bibnamefont {Rubeck}}, \bibinfo {author}
  {\bibfnamefont {B.}~\bibnamefont {Uchoa}}, \bibinfo {author} {\bibfnamefont
  {K.}~\bibnamefont {March}}, \bibinfo {author} {\bibfnamefont
  {C.}~\bibnamefont {Dwyer}}, \bibinfo {author} {\bibfnamefont
  {J.}~\bibnamefont {Schneeloch}}, \bibinfo {author} {\bibfnamefont
  {R.}~\bibnamefont {Zhong}}, \bibinfo {author} {\bibfnamefont {G.~D.}\
  \bibnamefont {Gu}}, \ and\ \bibinfo {author} {\bibfnamefont {P.}~\bibnamefont
  {Abbamonte}},\ }\href {\doibase 10.1103/PhysRevX.9.041062} {\bibfield
  {journal} {\bibinfo  {journal} {Phys. Rev. X}\ }\textbf {\bibinfo {volume}
  {9}},\ \bibinfo {pages} {041062} (\bibinfo {year} {2019})}\BibitemShut
  {NoStop}%
\bibitem [{\citenamefont {Arpaia}\ \emph {et~al.}(2023)\citenamefont {Arpaia},
  \citenamefont {Martinelli}, \citenamefont {Sala}, \citenamefont {Caprara},
  \citenamefont {Nag}, \citenamefont {Brookes}, \citenamefont {Camisa},
  \citenamefont {Li}, \citenamefont {Gao}, \citenamefont {Zhou}, \citenamefont
  {Garcia-Fernandez}, \citenamefont {Zhou}, \citenamefont {Schierle},
  \citenamefont {Bauch}, \citenamefont {Peng}, \citenamefont {Di~Castro},
  \citenamefont {Grilli}, \citenamefont {Lombardi}, \citenamefont
  {Braicovich},\ and\ \citenamefont {Ghiringhelli}}]{arpaia23}%
  \BibitemOpen
  \bibfield  {author} {\bibinfo {author} {\bibfnamefont {R.}~\bibnamefont
  {Arpaia}}, \bibinfo {author} {\bibfnamefont {L.}~\bibnamefont {Martinelli}},
  \bibinfo {author} {\bibfnamefont {M.~M.}\ \bibnamefont {Sala}}, \bibinfo
  {author} {\bibfnamefont {S.}~\bibnamefont {Caprara}}, \bibinfo {author}
  {\bibfnamefont {A.}~\bibnamefont {Nag}}, \bibinfo {author} {\bibfnamefont
  {N.~B.}\ \bibnamefont {Brookes}}, \bibinfo {author} {\bibfnamefont
  {P.}~\bibnamefont {Camisa}}, \bibinfo {author} {\bibfnamefont
  {Q.}~\bibnamefont {Li}}, \bibinfo {author} {\bibfnamefont {Q.}~\bibnamefont
  {Gao}}, \bibinfo {author} {\bibfnamefont {X.}~\bibnamefont {Zhou}}, \bibinfo
  {author} {\bibfnamefont {M.}~\bibnamefont {Garcia-Fernandez}}, \bibinfo
  {author} {\bibfnamefont {K.-J.}\ \bibnamefont {Zhou}}, \bibinfo {author}
  {\bibfnamefont {E.}~\bibnamefont {Schierle}}, \bibinfo {author}
  {\bibfnamefont {T.}~\bibnamefont {Bauch}}, \bibinfo {author} {\bibfnamefont
  {Y.~Y.}\ \bibnamefont {Peng}}, \bibinfo {author} {\bibfnamefont
  {C.}~\bibnamefont {Di~Castro}}, \bibinfo {author} {\bibfnamefont
  {M.}~\bibnamefont {Grilli}}, \bibinfo {author} {\bibfnamefont
  {F.}~\bibnamefont {Lombardi}}, \bibinfo {author} {\bibfnamefont
  {L.}~\bibnamefont {Braicovich}}, \ and\ \bibinfo {author} {\bibfnamefont
  {G.}~\bibnamefont {Ghiringhelli}},\ }\href {\doibase
  10.1038/s41467-023-42961-5} {\bibfield  {journal} {\bibinfo  {journal}
  {Nature Communications}\ }\textbf {\bibinfo {volume} {14}},\ \bibinfo {pages}
  {7198} (\bibinfo {year} {2023})}\BibitemShut {NoStop}%
\bibitem [{\citenamefont {Seibold}\ \emph {et~al.}(2021)\citenamefont
  {Seibold}, \citenamefont {Arpaia}, \citenamefont {Peng}, \citenamefont
  {Fumagalli}, \citenamefont {Braicovich}, \citenamefont {Di~Castro},
  \citenamefont {Grilli}, \citenamefont {Ghiringhelli},\ and\ \citenamefont
  {Caprara}}]{seibold21}%
  \BibitemOpen
  \bibfield  {author} {\bibinfo {author} {\bibfnamefont {G.}~\bibnamefont
  {Seibold}}, \bibinfo {author} {\bibfnamefont {R.}~\bibnamefont {Arpaia}},
  \bibinfo {author} {\bibfnamefont {Y.~Y.}\ \bibnamefont {Peng}}, \bibinfo
  {author} {\bibfnamefont {R.}~\bibnamefont {Fumagalli}}, \bibinfo {author}
  {\bibfnamefont {L.}~\bibnamefont {Braicovich}}, \bibinfo {author}
  {\bibfnamefont {C.}~\bibnamefont {Di~Castro}}, \bibinfo {author}
  {\bibfnamefont {M.}~\bibnamefont {Grilli}}, \bibinfo {author} {\bibfnamefont
  {G.~C.}\ \bibnamefont {Ghiringhelli}}, \ and\ \bibinfo {author}
  {\bibfnamefont {S.}~\bibnamefont {Caprara}},\ }\href {\doibase
  10.1038/s42005-020-00505-z} {\bibfield  {journal} {\bibinfo  {journal}
  {Communications Physics}\ }\textbf {\bibinfo {volume} {4}},\ \bibinfo {pages}
  {7} (\bibinfo {year} {2021})}\BibitemShut {NoStop}%
\bibitem [{\citenamefont {Caprara}\ \emph {et~al.}(2022)\citenamefont
  {Caprara}, \citenamefont {Castro}, \citenamefont {Mirarchi}, \citenamefont
  {Seibold},\ and\ \citenamefont {Grilli}}]{caprara22}%
  \BibitemOpen
  \bibfield  {author} {\bibinfo {author} {\bibfnamefont {S.}~\bibnamefont
  {Caprara}}, \bibinfo {author} {\bibfnamefont {C.~D.}\ \bibnamefont {Castro}},
  \bibinfo {author} {\bibfnamefont {G.}~\bibnamefont {Mirarchi}}, \bibinfo
  {author} {\bibfnamefont {G.}~\bibnamefont {Seibold}}, \ and\ \bibinfo
  {author} {\bibfnamefont {M.}~\bibnamefont {Grilli}},\ }\href {\doibase
  10.1038/s42005-021-00786-y} {\bibfield  {journal} {\bibinfo  {journal}
  {Communications Physics}\ }\textbf {\bibinfo {volume} {5}},\ \bibinfo {pages}
  {10} (\bibinfo {year} {2022})}\BibitemShut {NoStop}%
\bibitem [{\citenamefont {Greco}\ \emph {et~al.}(2019)\citenamefont {Greco},
  \citenamefont {Yamase},\ and\ \citenamefont {Bejas}}]{greco19}%
  \BibitemOpen
  \bibfield  {author} {\bibinfo {author} {\bibfnamefont {A.}~\bibnamefont
  {Greco}}, \bibinfo {author} {\bibfnamefont {H.}~\bibnamefont {Yamase}}, \
  and\ \bibinfo {author} {\bibfnamefont {M.}~\bibnamefont {Bejas}},\ }\href
  {\doibase 10.1038/s42005-018-0099-z} {\bibfield  {journal} {\bibinfo
  {journal} {Communications Physics}\ }\textbf {\bibinfo {volume} {2}},\
  \bibinfo {pages} {3} (\bibinfo {year} {2019})}\BibitemShut {NoStop}%
\bibitem [{\citenamefont {Greco}\ \emph {et~al.}(2020)\citenamefont {Greco},
  \citenamefont {Yamase},\ and\ \citenamefont {Bejas}}]{greco20}%
  \BibitemOpen
  \bibfield  {author} {\bibinfo {author} {\bibfnamefont {A.}~\bibnamefont
  {Greco}}, \bibinfo {author} {\bibfnamefont {H.}~\bibnamefont {Yamase}}, \
  and\ \bibinfo {author} {\bibfnamefont {M.}~\bibnamefont {Bejas}},\ }\href
  {\doibase 10.1103/PhysRevB.102.024509} {\bibfield  {journal} {\bibinfo
  {journal} {Phys. Rev. B}\ }\textbf {\bibinfo {volume} {102}},\ \bibinfo
  {pages} {024509} (\bibinfo {year} {2020})}\BibitemShut {NoStop}%
\bibitem [{\citenamefont {Yamase}\ \emph {et~al.}(2023)\citenamefont {Yamase},
  \citenamefont {Bejas},\ and\ \citenamefont {Greco}}]{yamase23}%
  \BibitemOpen
  \bibfield  {author} {\bibinfo {author} {\bibfnamefont {H.}~\bibnamefont
  {Yamase}}, \bibinfo {author} {\bibfnamefont {M.}~\bibnamefont {Bejas}}, \
  and\ \bibinfo {author} {\bibfnamefont {A.}~\bibnamefont {Greco}},\ }\href
  {\doibase 10.1038/s42005-023-01276-z} {\bibfield  {journal} {\bibinfo
  {journal} {Communications Physics}\ }\textbf {\bibinfo {volume} {6}},\
  \bibinfo {pages} {168} (\bibinfo {year} {2023})}\BibitemShut {NoStop}%
\bibitem [{\citenamefont {Yamase}\ \emph {et~al.}(2021)\citenamefont {Yamase},
  \citenamefont {Bejas},\ and\ \citenamefont {Greco}}]{yamase21a}%
  \BibitemOpen
  \bibfield  {author} {\bibinfo {author} {\bibfnamefont {H.}~\bibnamefont
  {Yamase}}, \bibinfo {author} {\bibfnamefont {M.}~\bibnamefont {Bejas}}, \
  and\ \bibinfo {author} {\bibfnamefont {A.}~\bibnamefont {Greco}},\ }\href
  {\doibase 10.1103/PhysRevB.104.045141} {\bibfield  {journal} {\bibinfo
  {journal} {Phys. Rev. B}\ }\textbf {\bibinfo {volume} {104}},\ \bibinfo
  {pages} {045141} (\bibinfo {year} {2021})}\BibitemShut {NoStop}%
\bibitem [{\citenamefont {Anderson}(1987)}]{anderson87}%
  \BibitemOpen
  \bibfield  {author} {\bibinfo {author} {\bibfnamefont {P.~W.}\ \bibnamefont
  {Anderson}},\ }\href {\doibase 10.1126/science.235.4793.1196} {\bibfield
  {journal} {\bibinfo  {journal} {Science}\ }\textbf {\bibinfo {volume}
  {235}},\ \bibinfo {pages} {1196} (\bibinfo {year} {1987})}\BibitemShut
  {NoStop}%
\bibitem [{\citenamefont {Zhang}\ and\ \citenamefont {Rice}(1988)}]{fczhang88}%
  \BibitemOpen
  \bibfield  {author} {\bibinfo {author} {\bibfnamefont {F.~C.}\ \bibnamefont
  {Zhang}}\ and\ \bibinfo {author} {\bibfnamefont {T.~M.}\ \bibnamefont
  {Rice}},\ }\href {\doibase 10.1103/PhysRevB.37.3759} {\bibfield  {journal}
  {\bibinfo  {journal} {Phys. Rev. B}\ }\textbf {\bibinfo {volume} {37}},\
  \bibinfo {pages} {3759} (\bibinfo {year} {1988})}\BibitemShut {NoStop}%
\bibitem [{\citenamefont {Lee}\ \emph {et~al.}(2006)\citenamefont {Lee},
  \citenamefont {Nagaosa},\ and\ \citenamefont {Wen}}]{lee06}%
  \BibitemOpen
  \bibfield  {author} {\bibinfo {author} {\bibfnamefont {P.~A.}\ \bibnamefont
  {Lee}}, \bibinfo {author} {\bibfnamefont {N.}~\bibnamefont {Nagaosa}}, \ and\
  \bibinfo {author} {\bibfnamefont {X.-G.}\ \bibnamefont {Wen}},\ }\href
  {\doibase 10.1103/RevModPhys.78.17} {\bibfield  {journal} {\bibinfo
  {journal} {Rev. Mod. Phys.}\ }\textbf {\bibinfo {volume} {78}},\ \bibinfo
  {pages} {17} (\bibinfo {year} {2006})}\BibitemShut {NoStop}%
\bibitem [{\citenamefont {Thio}\ \emph {et~al.}(1988)\citenamefont {Thio},
  \citenamefont {Thurston}, \citenamefont {Preyer}, \citenamefont {Picone},
  \citenamefont {Kastner}, \citenamefont {Jenssen}, \citenamefont {Gabbe},
  \citenamefont {Chen}, \citenamefont {Birgeneau},\ and\ \citenamefont
  {Aharony}}]{thio88}%
  \BibitemOpen
  \bibfield  {author} {\bibinfo {author} {\bibfnamefont {T.}~\bibnamefont
  {Thio}}, \bibinfo {author} {\bibfnamefont {T.~R.}\ \bibnamefont {Thurston}},
  \bibinfo {author} {\bibfnamefont {N.~W.}\ \bibnamefont {Preyer}}, \bibinfo
  {author} {\bibfnamefont {P.~J.}\ \bibnamefont {Picone}}, \bibinfo {author}
  {\bibfnamefont {M.~A.}\ \bibnamefont {Kastner}}, \bibinfo {author}
  {\bibfnamefont {H.~P.}\ \bibnamefont {Jenssen}}, \bibinfo {author}
  {\bibfnamefont {D.~R.}\ \bibnamefont {Gabbe}}, \bibinfo {author}
  {\bibfnamefont {C.~Y.}\ \bibnamefont {Chen}}, \bibinfo {author}
  {\bibfnamefont {R.~J.}\ \bibnamefont {Birgeneau}}, \ and\ \bibinfo {author}
  {\bibfnamefont {A.}~\bibnamefont {Aharony}},\ }\href {\doibase
  10.1103/PhysRevB.38.905} {\bibfield  {journal} {\bibinfo  {journal} {Phys.
  Rev. B}\ }\textbf {\bibinfo {volume} {38}},\ \bibinfo {pages} {905} (\bibinfo
  {year} {1988})}\BibitemShut {NoStop}%
\bibitem [{\citenamefont {Grecu}(1973)}]{grecu73}%
  \BibitemOpen
  \bibfield  {author} {\bibinfo {author} {\bibfnamefont {D.}~\bibnamefont
  {Grecu}},\ }\href {\doibase 10.1103/PhysRevB.8.1958} {\bibfield  {journal}
  {\bibinfo  {journal} {Phys. Rev. B}\ }\textbf {\bibinfo {volume} {8}},\
  \bibinfo {pages} {1958} (\bibinfo {year} {1973})}\BibitemShut {NoStop}%
\bibitem [{\citenamefont {Fetter}(1974)}]{fetter74}%
  \BibitemOpen
  \bibfield  {author} {\bibinfo {author} {\bibfnamefont {A.~L.}\ \bibnamefont
  {Fetter}},\ }\href {\doibase https://doi.org/10.1016/0003-4916(74)90397-2}
  {\bibfield  {journal} {\bibinfo  {journal} {Annals of Physics}\ }\textbf
  {\bibinfo {volume} {88}},\ \bibinfo {pages} {1} (\bibinfo {year}
  {1974})}\BibitemShut {NoStop}%
\bibitem [{\citenamefont {Grecu}(1975)}]{grecu75}%
  \BibitemOpen
  \bibfield  {author} {\bibinfo {author} {\bibfnamefont {D.}~\bibnamefont
  {Grecu}},\ }\href {\doibase 10.1088/0022-3719/8/16/014} {\bibfield  {journal}
  {\bibinfo  {journal} {J. Phys. C: Solid State Phys.}\ }\textbf {\bibinfo
  {volume} {8}},\ \bibinfo {pages} {2627} (\bibinfo {year} {1975})}\BibitemShut
  {NoStop}%
\bibitem [{\citenamefont {Becca}\ \emph {et~al.}(1996)\citenamefont {Becca},
  \citenamefont {Tarquini}, \citenamefont {Grilli},\ and\ \citenamefont
  {Di~Castro}}]{becca96}%
  \BibitemOpen
  \bibfield  {author} {\bibinfo {author} {\bibfnamefont {F.}~\bibnamefont
  {Becca}}, \bibinfo {author} {\bibfnamefont {M.}~\bibnamefont {Tarquini}},
  \bibinfo {author} {\bibfnamefont {M.}~\bibnamefont {Grilli}}, \ and\ \bibinfo
  {author} {\bibfnamefont {C.}~\bibnamefont {Di~Castro}},\ }\href {\doibase
  10.1103/PhysRevB.54.12443} {\bibfield  {journal} {\bibinfo  {journal} {Phys.
  Rev. B}\ }\textbf {\bibinfo {volume} {54}},\ \bibinfo {pages} {12443}
  (\bibinfo {year} {1996})}\BibitemShut {NoStop}%
\bibitem [{\citenamefont {Foussats}\ and\ \citenamefont
  {Greco}(2004)}]{foussats04}%
  \BibitemOpen
  \bibfield  {author} {\bibinfo {author} {\bibfnamefont {A.}~\bibnamefont
  {Foussats}}\ and\ \bibinfo {author} {\bibfnamefont {A.}~\bibnamefont
  {Greco}},\ }\href {\doibase 10.1103/PhysRevB.70.205123} {\bibfield  {journal}
  {\bibinfo  {journal} {Phys. Rev. B}\ }\textbf {\bibinfo {volume} {70}},\
  \bibinfo {pages} {205123} (\bibinfo {year} {2004})}\BibitemShut {NoStop}%
\bibitem [{\citenamefont {Bejas}\ \emph {et~al.}(2017)\citenamefont {Bejas},
  \citenamefont {Yamase},\ and\ \citenamefont {Greco}}]{bejas17}%
  \BibitemOpen
  \bibfield  {author} {\bibinfo {author} {\bibfnamefont {M.}~\bibnamefont
  {Bejas}}, \bibinfo {author} {\bibfnamefont {H.}~\bibnamefont {Yamase}}, \
  and\ \bibinfo {author} {\bibfnamefont {A.}~\bibnamefont {Greco}},\ }\href
  {\doibase 10.1103/PhysRevB.96.214513} {\bibfield  {journal} {\bibinfo
  {journal} {Phys. Rev. B}\ }\textbf {\bibinfo {volume} {96}},\ \bibinfo
  {pages} {214513} (\bibinfo {year} {2017})}\BibitemShut {NoStop}%
\bibitem [{\citenamefont {Varma}\ \emph {et~al.}(1989)\citenamefont {Varma},
  \citenamefont {Littlewood}, \citenamefont {Schmitt-Rink}, \citenamefont
  {Abrahams},\ and\ \citenamefont {Ruckenstein}}]{varma89}%
  \BibitemOpen
  \bibfield  {author} {\bibinfo {author} {\bibfnamefont {C.~M.}\ \bibnamefont
  {Varma}}, \bibinfo {author} {\bibfnamefont {P.~B.}\ \bibnamefont
  {Littlewood}}, \bibinfo {author} {\bibfnamefont {S.}~\bibnamefont
  {Schmitt-Rink}}, \bibinfo {author} {\bibfnamefont {E.}~\bibnamefont
  {Abrahams}}, \ and\ \bibinfo {author} {\bibfnamefont {A.~E.}\ \bibnamefont
  {Ruckenstein}},\ }\href {\doibase 10.1103/PhysRevLett.63.1996} {\bibfield
  {journal} {\bibinfo  {journal} {Phys. Rev. Lett.}\ }\textbf {\bibinfo
  {volume} {63}},\ \bibinfo {pages} {1996} (\bibinfo {year}
  {1989})}\BibitemShut {NoStop}%
\bibitem [{\citenamefont {Carbotte}\ \emph {et~al.}(2011)\citenamefont
  {Carbotte}, \citenamefont {Timusk},\ and\ \citenamefont
  {Hwang}}]{carbotte11}%
  \BibitemOpen
  \bibfield  {author} {\bibinfo {author} {\bibfnamefont {J.~P.}\ \bibnamefont
  {Carbotte}}, \bibinfo {author} {\bibfnamefont {T.}~\bibnamefont {Timusk}}, \
  and\ \bibinfo {author} {\bibfnamefont {J.}~\bibnamefont {Hwang}},\ }\href
  {\doibase 10.1088/0034-4885/74/6/066501} {\bibfield  {journal} {\bibinfo
  {journal} {Reports on Progress in Physics}\ }\textbf {\bibinfo {volume}
  {74}},\ \bibinfo {pages} {066501} (\bibinfo {year} {2011})}\BibitemShut
  {NoStop}%
\bibitem [{\citenamefont {Norman}\ \emph {et~al.}(2007)\citenamefont {Norman},
  \citenamefont {Kanigel}, \citenamefont {Randeria}, \citenamefont
  {Chatterjee},\ and\ \citenamefont {Campuzano}}]{norman07}%
  \BibitemOpen
  \bibfield  {author} {\bibinfo {author} {\bibfnamefont {M.~R.}\ \bibnamefont
  {Norman}}, \bibinfo {author} {\bibfnamefont {A.}~\bibnamefont {Kanigel}},
  \bibinfo {author} {\bibfnamefont {M.}~\bibnamefont {Randeria}}, \bibinfo
  {author} {\bibfnamefont {U.}~\bibnamefont {Chatterjee}}, \ and\ \bibinfo
  {author} {\bibfnamefont {J.~C.}\ \bibnamefont {Campuzano}},\ }\href {\doibase
  10.1103/PhysRevB.76.174501} {\bibfield  {journal} {\bibinfo  {journal} {Phys.
  Rev. B}\ }\textbf {\bibinfo {volume} {76}},\ \bibinfo {pages} {174501}
  (\bibinfo {year} {2007})}\BibitemShut {NoStop}%
\bibitem [{\citenamefont {K\"uspert}\ \emph {et~al.}(2022)\citenamefont
  {K\"uspert}, \citenamefont {Cohn~Wagner}, \citenamefont {Lin}, \citenamefont
  {von Arx}, \citenamefont {Wang}, \citenamefont {Kramer}, \citenamefont
  {Pudelko}, \citenamefont {Plumb}, \citenamefont {Matt}, \citenamefont
  {Fatuzzo}, \citenamefont {Sutter}, \citenamefont {Sassa}, \citenamefont
  {Yan}, \citenamefont {Zhou}, \citenamefont {Goodenough}, \citenamefont
  {Pyon}, \citenamefont {Takayama}, \citenamefont {Takagi}, \citenamefont
  {Kurosawa}, \citenamefont {Momono}, \citenamefont {Oda}, \citenamefont
  {Hoesch}, \citenamefont {Cacho}, \citenamefont {Kim}, \citenamefont {Horio},\
  and\ \citenamefont {Chang}}]{kuspert22}%
  \BibitemOpen
  \bibfield  {author} {\bibinfo {author} {\bibfnamefont {J.}~\bibnamefont
  {K\"uspert}}, \bibinfo {author} {\bibfnamefont {R.}~\bibnamefont
  {Cohn~Wagner}}, \bibinfo {author} {\bibfnamefont {C.}~\bibnamefont {Lin}},
  \bibinfo {author} {\bibfnamefont {K.}~\bibnamefont {von Arx}}, \bibinfo
  {author} {\bibfnamefont {Q.}~\bibnamefont {Wang}}, \bibinfo {author}
  {\bibfnamefont {K.}~\bibnamefont {Kramer}}, \bibinfo {author} {\bibfnamefont
  {W.~R.}\ \bibnamefont {Pudelko}}, \bibinfo {author} {\bibfnamefont {N.~C.}\
  \bibnamefont {Plumb}}, \bibinfo {author} {\bibfnamefont {C.~E.}\ \bibnamefont
  {Matt}}, \bibinfo {author} {\bibfnamefont {C.~G.}\ \bibnamefont {Fatuzzo}},
  \bibinfo {author} {\bibfnamefont {D.}~\bibnamefont {Sutter}}, \bibinfo
  {author} {\bibfnamefont {Y.}~\bibnamefont {Sassa}}, \bibinfo {author}
  {\bibfnamefont {J.-Q.}\ \bibnamefont {Yan}}, \bibinfo {author} {\bibfnamefont
  {J.-S.}\ \bibnamefont {Zhou}}, \bibinfo {author} {\bibfnamefont {J.~B.}\
  \bibnamefont {Goodenough}}, \bibinfo {author} {\bibfnamefont
  {S.}~\bibnamefont {Pyon}}, \bibinfo {author} {\bibfnamefont {T.}~\bibnamefont
  {Takayama}}, \bibinfo {author} {\bibfnamefont {H.}~\bibnamefont {Takagi}},
  \bibinfo {author} {\bibfnamefont {T.}~\bibnamefont {Kurosawa}}, \bibinfo
  {author} {\bibfnamefont {N.}~\bibnamefont {Momono}}, \bibinfo {author}
  {\bibfnamefont {M.}~\bibnamefont {Oda}}, \bibinfo {author} {\bibfnamefont
  {M.}~\bibnamefont {Hoesch}}, \bibinfo {author} {\bibfnamefont
  {C.}~\bibnamefont {Cacho}}, \bibinfo {author} {\bibfnamefont {T.~K.}\
  \bibnamefont {Kim}}, \bibinfo {author} {\bibfnamefont {M.}~\bibnamefont
  {Horio}}, \ and\ \bibinfo {author} {\bibfnamefont {J.}~\bibnamefont
  {Chang}},\ }\href {\doibase 10.1103/PhysRevResearch.4.043015} {\bibfield
  {journal} {\bibinfo  {journal} {Phys. Rev. Res.}\ }\textbf {\bibinfo {volume}
  {4}},\ \bibinfo {pages} {043015} (\bibinfo {year} {2022})}\BibitemShut
  {NoStop}%
\bibitem [{mis()}]{misc-factor2a}%
  \BibitemOpen
  \href@noop {} {}\bibinfo {note} {The factor of 1/2 originates from the
  large-$N$ formalism and $N=2$ corresponds to a physical value.}\BibitemShut
  {Stop}%
\bibitem [{\citenamefont {Norman}\ \emph {et~al.}(1998)\citenamefont {Norman},
  \citenamefont {Randeria}, \citenamefont {Ding},\ and\ \citenamefont
  {Campuzano}}]{norman98a}%
  \BibitemOpen
  \bibfield  {author} {\bibinfo {author} {\bibfnamefont {M.~R.}\ \bibnamefont
  {Norman}}, \bibinfo {author} {\bibfnamefont {M.}~\bibnamefont {Randeria}},
  \bibinfo {author} {\bibfnamefont {H.}~\bibnamefont {Ding}}, \ and\ \bibinfo
  {author} {\bibfnamefont {J.~C.}\ \bibnamefont {Campuzano}},\ }\href {\doibase
  10.1103/PhysRevB.57.R11093} {\bibfield  {journal} {\bibinfo  {journal} {Phys.
  Rev. B}\ }\textbf {\bibinfo {volume} {57}},\ \bibinfo {pages} {R11093}
  (\bibinfo {year} {1998})}\BibitemShut {NoStop}%
\bibitem [{\citenamefont {Kanigel}\ \emph {et~al.}(2007)\citenamefont
  {Kanigel}, \citenamefont {Chatterjee}, \citenamefont {Randeria},
  \citenamefont {Norman}, \citenamefont {Souma}, \citenamefont {Shi},
  \citenamefont {Li}, \citenamefont {Raffy},\ and\ \citenamefont
  {Campuzano}}]{kanigel07}%
  \BibitemOpen
  \bibfield  {author} {\bibinfo {author} {\bibfnamefont {A.}~\bibnamefont
  {Kanigel}}, \bibinfo {author} {\bibfnamefont {U.}~\bibnamefont {Chatterjee}},
  \bibinfo {author} {\bibfnamefont {M.}~\bibnamefont {Randeria}}, \bibinfo
  {author} {\bibfnamefont {M.~R.}\ \bibnamefont {Norman}}, \bibinfo {author}
  {\bibfnamefont {S.}~\bibnamefont {Souma}}, \bibinfo {author} {\bibfnamefont
  {M.}~\bibnamefont {Shi}}, \bibinfo {author} {\bibfnamefont {Z.~Z.}\
  \bibnamefont {Li}}, \bibinfo {author} {\bibfnamefont {H.}~\bibnamefont
  {Raffy}}, \ and\ \bibinfo {author} {\bibfnamefont {J.~C.}\ \bibnamefont
  {Campuzano}},\ }\href {\doibase 10.1103/PhysRevLett.99.157001} {\bibfield
  {journal} {\bibinfo  {journal} {Phys. Rev. Lett.}\ }\textbf {\bibinfo
  {volume} {99}},\ \bibinfo {pages} {157001} (\bibinfo {year}
  {2007})}\BibitemShut {NoStop}%
\bibitem [{\citenamefont {Abdel-Jawad}\ \emph {et~al.}(2006)\citenamefont
  {Abdel-Jawad}, \citenamefont {Kennett}, \citenamefont {Balicas},
  \citenamefont {Carrington}, \citenamefont {Mackenzie}, \citenamefont
  {McKenzie},\ and\ \citenamefont {Hussey}}]{jawad06}%
  \BibitemOpen
  \bibfield  {author} {\bibinfo {author} {\bibfnamefont {M.}~\bibnamefont
  {Abdel-Jawad}}, \bibinfo {author} {\bibfnamefont {M.~P.}\ \bibnamefont
  {Kennett}}, \bibinfo {author} {\bibfnamefont {L.}~\bibnamefont {Balicas}},
  \bibinfo {author} {\bibfnamefont {A.}~\bibnamefont {Carrington}}, \bibinfo
  {author} {\bibfnamefont {A.~P.}\ \bibnamefont {Mackenzie}}, \bibinfo {author}
  {\bibfnamefont {R.~H.}\ \bibnamefont {McKenzie}}, \ and\ \bibinfo {author}
  {\bibfnamefont {N.~E.}\ \bibnamefont {Hussey}},\ }\href {\doibase
  10.1038/nphys449} {\bibfield  {journal} {\bibinfo  {journal} {Nature
  Physics}\ }\textbf {\bibinfo {volume} {2}},\ \bibinfo {pages} {821} (\bibinfo
  {year} {2006})}\BibitemShut {NoStop}%
\bibitem [{\citenamefont {Abdel-Jawad}\ \emph {et~al.}(2007)\citenamefont
  {Abdel-Jawad}, \citenamefont {Analytis}, \citenamefont {Balicas},
  \citenamefont {Carrington}, \citenamefont {Charmant}, \citenamefont
  {French},\ and\ \citenamefont {Hussey}}]{jawad07}%
  \BibitemOpen
  \bibfield  {author} {\bibinfo {author} {\bibfnamefont {M.}~\bibnamefont
  {Abdel-Jawad}}, \bibinfo {author} {\bibfnamefont {J.~G.}\ \bibnamefont
  {Analytis}}, \bibinfo {author} {\bibfnamefont {L.}~\bibnamefont {Balicas}},
  \bibinfo {author} {\bibfnamefont {A.}~\bibnamefont {Carrington}}, \bibinfo
  {author} {\bibfnamefont {J.~P.~H.}\ \bibnamefont {Charmant}}, \bibinfo
  {author} {\bibfnamefont {M.~M.~J.}\ \bibnamefont {French}}, \ and\ \bibinfo
  {author} {\bibfnamefont {N.~E.}\ \bibnamefont {Hussey}},\ }\href {\doibase
  10.1103/PhysRevLett.99.107002} {\bibfield  {journal} {\bibinfo  {journal}
  {Phys. Rev. Lett.}\ }\textbf {\bibinfo {volume} {99}},\ \bibinfo {pages}
  {107002} (\bibinfo {year} {2007})}\BibitemShut {NoStop}%
\bibitem [{\citenamefont {French}\ \emph {et~al.}(2009)\citenamefont {French},
  \citenamefont {Analytis}, \citenamefont {Carrington}, \citenamefont
  {Balicas},\ and\ \citenamefont {Hussey}}]{french09}%
  \BibitemOpen
  \bibfield  {author} {\bibinfo {author} {\bibfnamefont {M.~M.~J.}\
  \bibnamefont {French}}, \bibinfo {author} {\bibfnamefont {J.~G.}\
  \bibnamefont {Analytis}}, \bibinfo {author} {\bibfnamefont {A.}~\bibnamefont
  {Carrington}}, \bibinfo {author} {\bibfnamefont {L.}~\bibnamefont {Balicas}},
  \ and\ \bibinfo {author} {\bibfnamefont {N.~E.}\ \bibnamefont {Hussey}},\
  }\href {\doibase 10.1088/1367-2630/11/5/055057} {\bibfield  {journal}
  {\bibinfo  {journal} {New Journal of Physics}\ }\textbf {\bibinfo {volume}
  {11}},\ \bibinfo {pages} {055057} (\bibinfo {year} {2009})}\BibitemShut
  {NoStop}%
\bibitem [{\citenamefont {Nakamae}\ \emph {et~al.}(2003)\citenamefont
  {Nakamae}, \citenamefont {Behnia}, \citenamefont {Mangkorntong},
  \citenamefont {Nohara}, \citenamefont {Takagi}, \citenamefont {Yates},\ and\
  \citenamefont {Hussey}}]{nakamae03}%
  \BibitemOpen
  \bibfield  {author} {\bibinfo {author} {\bibfnamefont {S.}~\bibnamefont
  {Nakamae}}, \bibinfo {author} {\bibfnamefont {K.}~\bibnamefont {Behnia}},
  \bibinfo {author} {\bibfnamefont {N.}~\bibnamefont {Mangkorntong}}, \bibinfo
  {author} {\bibfnamefont {M.}~\bibnamefont {Nohara}}, \bibinfo {author}
  {\bibfnamefont {H.}~\bibnamefont {Takagi}}, \bibinfo {author} {\bibfnamefont
  {S.~J.~C.}\ \bibnamefont {Yates}}, \ and\ \bibinfo {author} {\bibfnamefont
  {N.~E.}\ \bibnamefont {Hussey}},\ }\href {\doibase
  10.1103/PhysRevB.68.100502} {\bibfield  {journal} {\bibinfo  {journal} {Phys.
  Rev. B}\ }\textbf {\bibinfo {volume} {68}},\ \bibinfo {pages} {100502}
  (\bibinfo {year} {2003})}\BibitemShut {NoStop}%
\bibitem [{\citenamefont {Cooper}\ \emph {et~al.}(2009)\citenamefont {Cooper},
  \citenamefont {Wang}, \citenamefont {Vignolle}, \citenamefont {Lipscombe},
  \citenamefont {Hayden}, \citenamefont {Tanabe}, \citenamefont {Adachi},
  \citenamefont {Koike}, \citenamefont {Nohara}, \citenamefont {Takagi},
  \citenamefont {Proust},\ and\ \citenamefont {Hussey}}]{cooper09}%
  \BibitemOpen
  \bibfield  {author} {\bibinfo {author} {\bibfnamefont {R.~A.}\ \bibnamefont
  {Cooper}}, \bibinfo {author} {\bibfnamefont {Y.}~\bibnamefont {Wang}},
  \bibinfo {author} {\bibfnamefont {B.}~\bibnamefont {Vignolle}}, \bibinfo
  {author} {\bibfnamefont {O.~J.}\ \bibnamefont {Lipscombe}}, \bibinfo {author}
  {\bibfnamefont {S.~M.}\ \bibnamefont {Hayden}}, \bibinfo {author}
  {\bibfnamefont {Y.}~\bibnamefont {Tanabe}}, \bibinfo {author} {\bibfnamefont
  {T.}~\bibnamefont {Adachi}}, \bibinfo {author} {\bibfnamefont
  {Y.}~\bibnamefont {Koike}}, \bibinfo {author} {\bibfnamefont
  {M.}~\bibnamefont {Nohara}}, \bibinfo {author} {\bibfnamefont
  {H.}~\bibnamefont {Takagi}}, \bibinfo {author} {\bibfnamefont
  {C.}~\bibnamefont {Proust}}, \ and\ \bibinfo {author} {\bibfnamefont {N.~E.}\
  \bibnamefont {Hussey}},\ }\href {\doibase 10.1126/science.1165015} {\bibfield
   {journal} {\bibinfo  {journal} {Science}\ }\textbf {\bibinfo {volume}
  {323}},\ \bibinfo {pages} {603} (\bibinfo {year} {2009})}\BibitemShut
  {NoStop}%
\bibitem [{\citenamefont {Harada}\ \emph {et~al.}(2022)\citenamefont {Harada},
  \citenamefont {Teramoto}, \citenamefont {Usui}, \citenamefont {Itaka},
  \citenamefont {Fujii}, \citenamefont {Noji}, \citenamefont {Taniguchi},
  \citenamefont {Matsukawa}, \citenamefont {Ishikawa}, \citenamefont {Kindo},
  \citenamefont {Dessau},\ and\ \citenamefont {Watanabe}}]{harada22}%
  \BibitemOpen
  \bibfield  {author} {\bibinfo {author} {\bibfnamefont {K.}~\bibnamefont
  {Harada}}, \bibinfo {author} {\bibfnamefont {Y.}~\bibnamefont {Teramoto}},
  \bibinfo {author} {\bibfnamefont {T.}~\bibnamefont {Usui}}, \bibinfo {author}
  {\bibfnamefont {K.}~\bibnamefont {Itaka}}, \bibinfo {author} {\bibfnamefont
  {T.}~\bibnamefont {Fujii}}, \bibinfo {author} {\bibfnamefont
  {T.}~\bibnamefont {Noji}}, \bibinfo {author} {\bibfnamefont {H.}~\bibnamefont
  {Taniguchi}}, \bibinfo {author} {\bibfnamefont {M.}~\bibnamefont
  {Matsukawa}}, \bibinfo {author} {\bibfnamefont {H.}~\bibnamefont {Ishikawa}},
  \bibinfo {author} {\bibfnamefont {K.}~\bibnamefont {Kindo}}, \bibinfo
  {author} {\bibfnamefont {D.~S.}\ \bibnamefont {Dessau}}, \ and\ \bibinfo
  {author} {\bibfnamefont {T.}~\bibnamefont {Watanabe}},\ }\href {\doibase
  10.1103/PhysRevB.105.085131} {\bibfield  {journal} {\bibinfo  {journal}
  {Phys. Rev. B}\ }\textbf {\bibinfo {volume} {105}},\ \bibinfo {pages}
  {085131} (\bibinfo {year} {2022})}\BibitemShut {NoStop}%
\bibitem [{\citenamefont {Takagi}\ \emph {et~al.}(1992)\citenamefont {Takagi},
  \citenamefont {Batlogg}, \citenamefont {Kao}, \citenamefont {Kwo},
  \citenamefont {Cava}, \citenamefont {Krajewski},\ and\ \citenamefont
  {Peck}}]{takagi92}%
  \BibitemOpen
  \bibfield  {author} {\bibinfo {author} {\bibfnamefont {H.}~\bibnamefont
  {Takagi}}, \bibinfo {author} {\bibfnamefont {B.}~\bibnamefont {Batlogg}},
  \bibinfo {author} {\bibfnamefont {H.~L.}\ \bibnamefont {Kao}}, \bibinfo
  {author} {\bibfnamefont {J.}~\bibnamefont {Kwo}}, \bibinfo {author}
  {\bibfnamefont {R.~J.}\ \bibnamefont {Cava}}, \bibinfo {author}
  {\bibfnamefont {J.~J.}\ \bibnamefont {Krajewski}}, \ and\ \bibinfo {author}
  {\bibfnamefont {W.~F.}\ \bibnamefont {Peck}},\ }\href {\doibase
  10.1103/PhysRevLett.69.2975} {\bibfield  {journal} {\bibinfo  {journal}
  {Phys. Rev. Lett.}\ }\textbf {\bibinfo {volume} {69}},\ \bibinfo {pages}
  {2975} (\bibinfo {year} {1992})}\BibitemShut {NoStop}%
\bibitem [{\citenamefont {Vignolle}\ \emph {et~al.}(2008)\citenamefont
  {Vignolle}, \citenamefont {Carrington}, \citenamefont {Cooper}, \citenamefont
  {French}, \citenamefont {Mackenzie}, \citenamefont {Jaudet}, \citenamefont
  {Vignolles}, \citenamefont {Proust},\ and\ \citenamefont
  {Hussey}}]{vignolle08}%
  \BibitemOpen
  \bibfield  {author} {\bibinfo {author} {\bibfnamefont {B.}~\bibnamefont
  {Vignolle}}, \bibinfo {author} {\bibfnamefont {A.}~\bibnamefont
  {Carrington}}, \bibinfo {author} {\bibfnamefont {R.~A.}\ \bibnamefont
  {Cooper}}, \bibinfo {author} {\bibfnamefont {M.~M.~J.}\ \bibnamefont
  {French}}, \bibinfo {author} {\bibfnamefont {A.~P.}\ \bibnamefont
  {Mackenzie}}, \bibinfo {author} {\bibfnamefont {C.}~\bibnamefont {Jaudet}},
  \bibinfo {author} {\bibfnamefont {D.}~\bibnamefont {Vignolles}}, \bibinfo
  {author} {\bibfnamefont {C.}~\bibnamefont {Proust}}, \ and\ \bibinfo {author}
  {\bibfnamefont {N.~E.}\ \bibnamefont {Hussey}},\ }\href {\doibase
  10.1038/nature07323} {\bibfield  {journal} {\bibinfo  {journal} {Nature}\
  }\textbf {\bibinfo {volume} {455}},\ \bibinfo {pages} {952} (\bibinfo {year}
  {2008})}\BibitemShut {NoStop}%
\bibitem [{\citenamefont {Johnson}\ \emph {et~al.}(2001)\citenamefont
  {Johnson}, \citenamefont {Valla}, \citenamefont {Fedorov}, \citenamefont
  {Yusof}, \citenamefont {Wells}, \citenamefont {Li}, \citenamefont
  {Moodenbaugh}, \citenamefont {Gu}, \citenamefont {Koshizuka}, \citenamefont
  {Kendziora}, \citenamefont {Jian},\ and\ \citenamefont {Hinks}}]{johnson01}%
  \BibitemOpen
  \bibfield  {author} {\bibinfo {author} {\bibfnamefont {P.~D.}\ \bibnamefont
  {Johnson}}, \bibinfo {author} {\bibfnamefont {T.}~\bibnamefont {Valla}},
  \bibinfo {author} {\bibfnamefont {A.~V.}\ \bibnamefont {Fedorov}}, \bibinfo
  {author} {\bibfnamefont {Z.}~\bibnamefont {Yusof}}, \bibinfo {author}
  {\bibfnamefont {B.~O.}\ \bibnamefont {Wells}}, \bibinfo {author}
  {\bibfnamefont {Q.}~\bibnamefont {Li}}, \bibinfo {author} {\bibfnamefont
  {A.~R.}\ \bibnamefont {Moodenbaugh}}, \bibinfo {author} {\bibfnamefont
  {G.~D.}\ \bibnamefont {Gu}}, \bibinfo {author} {\bibfnamefont
  {N.}~\bibnamefont {Koshizuka}}, \bibinfo {author} {\bibfnamefont
  {C.}~\bibnamefont {Kendziora}}, \bibinfo {author} {\bibfnamefont
  {S.}~\bibnamefont {Jian}}, \ and\ \bibinfo {author} {\bibfnamefont {D.~G.}\
  \bibnamefont {Hinks}},\ }\href {\doibase 10.1103/PhysRevLett.87.177007}
  {\bibfield  {journal} {\bibinfo  {journal} {Phys. Rev. Lett.}\ }\textbf
  {\bibinfo {volume} {87}},\ \bibinfo {pages} {177007} (\bibinfo {year}
  {2001})}\BibitemShut {NoStop}%
\bibitem [{\citenamefont {Giuliani}\ and\ \citenamefont
  {Vignale}(2005)}]{giuliani}%
  \BibitemOpen
  \bibfield  {author} {\bibinfo {author} {\bibfnamefont {G.~F.}\ \bibnamefont
  {Giuliani}}\ and\ \bibinfo {author} {\bibfnamefont {G.}~\bibnamefont
  {Vignale}},\ }\href@noop {} {\emph {\bibinfo {title} {Quantum Theory of the
  Electron Liquid}}}\ (\bibinfo  {publisher} {Cambridge University Press},\
  \bibinfo {year} {2005})\BibitemShut {NoStop}%
\bibitem [{\citenamefont {Yang}\ \emph {et~al.}(2006)\citenamefont {Yang},
  \citenamefont {Rice},\ and\ \citenamefont {Zhang}}]{yang06}%
  \BibitemOpen
  \bibfield  {author} {\bibinfo {author} {\bibfnamefont {K.-Y.}\ \bibnamefont
  {Yang}}, \bibinfo {author} {\bibfnamefont {T.~M.}\ \bibnamefont {Rice}}, \
  and\ \bibinfo {author} {\bibfnamefont {F.-C.}\ \bibnamefont {Zhang}},\ }\href
  {\doibase 10.1103/PhysRevB.73.174501} {\bibfield  {journal} {\bibinfo
  {journal} {Phys. Rev. B}\ }\textbf {\bibinfo {volume} {73}},\ \bibinfo
  {pages} {174501} (\bibinfo {year} {2006})}\BibitemShut {NoStop}%
\bibitem [{\citenamefont {Yamase}\ and\ \citenamefont
  {Metzner}(2012)}]{yamase12}%
  \BibitemOpen
  \bibfield  {author} {\bibinfo {author} {\bibfnamefont {H.}~\bibnamefont
  {Yamase}}\ and\ \bibinfo {author} {\bibfnamefont {W.}~\bibnamefont
  {Metzner}},\ }\href {\doibase 10.1103/PhysRevLett.108.186405} {\bibfield
  {journal} {\bibinfo  {journal} {Phys. Rev. Lett.}\ }\textbf {\bibinfo
  {volume} {108}},\ \bibinfo {pages} {186405} (\bibinfo {year}
  {2012})}\BibitemShut {NoStop}%
\end{thebibliography}%

\end{document}